\documentclass{aa}
\usepackage{graphicx}
\usepackage{txfonts}
\usepackage{natbib}
\begin{document}
   \title{The infrared emission of  ultraviolet selected galaxies from $z = 0$ to $z=1$}


   \author{V.\ Buat\inst{1}\and
          T.\ T.\ Takeuchi\inst{2}\and
          D.\ Burgarella\inst{1}\and
          E.\ Giovannnoli\inst{1}\and
          K.\ L.\ Murata\inst{3}}

   \offprints{V. Buat}

   \institute{Laboratoire d'Astrophysique de Marseille, OAMP, Universit\'e Aix-marseille, CNRS, 38 rue Fr\'ed\'eric Joliot-Curie, 13388 Marseille cedex 13, France\\
              \email{veronique.buat@oamp.fr,denis.burgarella@oamp.fr,elodie.giovannoli@oamp.fr}
         \and
             Institute for Advanced Research, Nagoya University\\
             \email{takeuchi@iar.nagoya-u.ac.jp}
         \and
             Division of Particle and Astrophysical Sciences,  
             Nagoya University\\
             \email{murata.katsuhiro@g.mbox.nagoya-u.ac.jp}
                         }

   \date{}

 
  \abstract
   {}
   { We want to study  the IR ($> 8\;\mu$m) emission of samples of  galaxies selected in their rest-frame UV in a very homogeneous way (wavelength and luminosity) from  $z = 0$ to $z=1$: comparing their UV and IR rest-frame emissions will allow us to study the evolution of dust attenuation with $z$ as well as to check if a UV selection is able to track all the star formation. This UV selection will also be compared to a sample of Lyman Break Galaxies selected at $z \simeq 1$.}
   {We select galaxies in UV (1500--1800~\AA) rest-frame at  $z=0$, $z=0.6\mbox{--}0.8$ and $z=0.8\mbox{--}1.2$ together with a sample of Lyman Break Galaxies at $z=0.9\mbox{--}1.3$, the samples are  built  in order to sample the same range of luminosity at any redshift. The UV rest-frame data come from GALEX  for $z<1$ and the $U$-band of the EIS survey (at $z=1$). The UV data are combined to the IRAS  60~$\mu$m observations at $z=0$ and the SPITZER data at $24\;\mu$m for $z>0$ sources. The evolution of the IR and UV luminosities  with $z$ is analysed for individual galaxies as well as in terms of luminosity functions.}
   {The $L_{\rm IR}/L_{\rm UV}$ ratio is used to measure dust attenuation. This ratio does not exhibit a strong evolution with $z$ for the bulk of our sample galaxies but some trends are found for galaxies  with a strong dust attenuation and  for UV luminous sources: galaxies with $L_{\rm IR}/L_{\rm UV}>10$  are more frequent at $z>0$ than at $z=0$ and the largest values of $L_{\rm IR}/L_{\rm UV}$ are found for UV faint objects; conversely the most  luminous   galaxies of our samples ($L_{\rm UV}> 2 \times 10^{10} L_\odot$), detected at $z=1$,  exhibit a lower dust attenuation than the fainter ones. $L_{\rm IR}/L_{\rm UV}$ increases with the $K$ rest-frame luminosity of the galaxies at all the redshifts considered and shows a residual anti-correlation with $L_{\rm UV}$.  The most massive and UV luminous galaxies exhibit quite large  specific star formation rates.  Lyman Break Galaxies exhibit systematically  lower dust attenuation than UV selected galaxies of same luminosity but similar specific star formation rates. 
   
The analysis of the UV+IR luminosity functions leads to the conclusion that up to $z = 1$ most of the star formation activity of UV selected galaxies is emitted in IR. Whereas we are able to retrieve all the star formation from our  UV selection at $z=0.7$, at $z = 1$  we miss a large fraction of galaxies more luminous than $\simeq 10^{11} L_{\odot}$. The effect is found larger for Lyman Break Galaxies.}
{}
\titlerunning{UV selected galaxies from $z = 0$ to 1}
   \maketitle
%

\section{Introduction}

The measure of the star formation rate at various redshifts is one of the most spread out diagnostic  to quantify the evolution of the galaxies, individually or for the population as a whole. The best way to perform such an analysis is to rely on galaxy samples selected in the same way at different redshift. As long as star forming galaxies have to be selected a UV selection is theoretically very efficient. Nevertheless, the situation is made difficult because of dust attenuation. Indeed dust attenuation affects the measure of the star formation rate derived from the observed UV emission and a correction (sometimes quite large) must be applied to the observed UV emission before translating it into SFR. However the effects of dust obscuration may be even more dramatic if they lead to a loss of galaxies which might  not be detected at all in UV surveys: in such a case  a correction for dust attenuation of the total light observed in UV would not be sufficient to retrieve the whole star formation at a given redshift.\\

The UV (1500--1800~\AA) to IR (8--1000~$\mu$m)  luminosity ratio $L_{\rm IR}/L_{\rm UV}$ is now commonly used as a robust proxy for  dust attenuation. The GALEX all sky survey associated with the IRAS catalogues has produced large samples of nearby galaxies observed at both wavelengths and used to study the variation of $L_{\rm IR}/L_{\rm UV}$ with the total $L_{\rm IR}+L_{\rm UV}$ luminosity (i.e. the total SFR of galaxies). $L_{\rm IR}/L_{\rm UV}$
 has been found to increase with $L_{\rm IR}+L_{\rm UV}$ \citep{martin05,buat07-1}. The same relation holds for galaxies selected either in UV or in IR \citep{buat07-1}.
This confirms the  general increase of dust attenuation with the luminosity of the galaxies already  reported   \citep[e.g.][]{hopkins01,moustakas06}. \\
Such a relation between total luminosity and dust attenuation in galaxies is useful  to correct for systematic effects of dust attenuation in galaxy surveys. It is important to check if this relation still holds at higher  $z$. Several studies have been devoted to this issue. Contrary to what is found in the nearby universe, it appears that the result depends on the way galaxies are selected. IR selected galaxies at intermediate redshift ($z=0.5\mbox{--}0.8$) seem to follow the mean trends found at $z = 0$ between $L_{\rm IR}+L_{\rm UV}$ and $L_{\rm IR}/L_{\rm UV}$ \citep{choi06,xu06,zheng07}. \citet{buat07-2} reported only a slight decrease of dust attenuation ($\sim 0.5$ mag) for luminous IR galaxies (LIRGs) at $z = 0.7$ as compared to a similar sample of galaxies at $z=0$. 
When galaxies are selected in UV/optical the situation is quite different: it has been reported    a strong decrease of 
$L_{\rm IR}/L_{\rm UV}$ for a given total $L_{\rm IR}+L_{\rm UV}$ as compared to  what is 
found at $z = 0$ \citep{burgarella06,burgarella07,reddy08}.\\
The origin of this discrepancy is not clear.  Are the properties of galaxies changing with $z$ and/or do we sample 
very different galaxy populations when selecting in UV or in IR? 
In the nearby universe \citet{buat07-1} have shown that a UV or an IR selection leads to similar results except for 
intrinsically very luminous galaxies which are under-represented in the UV samples. Such objects are 
rare at $z = 0$ but  because of  the evolution of the luminosity functions with $z$ we might expect a different 
situation at higher $z$.\\
In order to answer to these questions, we gather here several samples of UV selected galaxies from 
$z=0$ to $z=1.2$ selected in a very homogeneous way. 
We  also consider a sample  of Lyman Break Galaxies (LBGs) selected at 
$z \simeq 1$ to be compared to our pure UV selections.  
We  add IR fluxes (when available) for all the galaxies of these samples.  With these data  we want to study the evolution with  $z$ of the IR emission  mainly through the  study of the  $L_{\rm IR}/L_{\rm UV}$  ratio and to measure the total star formation activity by combining IR and UV emissions. The reliability of UV selected galaxies to trace all the star formation  will be discussed through the analysis of bolometric ($L_{\rm IR}+L_{\rm UV}$) luminosity functions.\\

 Throughout the paper we will assume $\Omega_m = 0.27$, $\Omega_{\Lambda} = 0.73$ and $ H_0 = 71 {\rm~ km~ s^{-1}~ Mpc^{-1}}$. All the magnitudes are given in the AB system except for the R magnitude from the COMBO-17 survey (section 2.1). The luminosities are defined as $\nu L_{\nu}$ and expressed  in solar units  assuming  
$L_{\odot} = 3.83 \times 10^{33} {\rm ~erg~ s^{-1}}$

\section{The galaxy samples}

\subsection{The UV selected samples}

The samples must be  purely UV selected. As a consequence we mostly rely on the GALEX survey for 
redshift lower than 1. \\
At $z=0$, we take the IRAS/GALEX sample  built by \citet{buat07-1}. 
This sample consists of galaxies selected in the GALEX FUV band (1530~\AA) and with $ \rm FUV < 17 mag$.  
It is a flux limited sample and  the luminosity function has  been built down to $L_{\rm FUV}= 10^8 L_{\odot}$. 
The FUV wavelength will be taken as the reference wavelength for the samples at higher $z$, 
and quoted as UV throughout the paper. 

At higher $z$, our samples are extracted from the GALEX deep observations of the CDFS.
GALEX \citep{morissey05} observed this field in both the FUV
(1530~\AA) and the NUV (2310~\AA) as part of its deep imaging
survey.  In order to add IR data to the galaxy sample we restrict our study to the sub-field covered by 
SPITZER/MIPS  observations as part of the GOODS key program  \citep[e.g.][]{elbaz07}. 

The sample at $z \simeq 0.7$  was  already used by \citet{buat08}, at this redshift  the NUV band of 
GALEX at 2310~\AA\ corresponds to the FUV rest frame of the galaxies.
This sample is thus made of galaxies selected in NUV and with a redshift comprised between 0.6 and 0.8. 
Redshifts come from the COMBO-17 survey of the field \citep{wolf04}. The reduction of the data and the 
cross-identification with the COMBO-17
sources is also described in \citet{burgarella06}.
For the brightest isolated sources of the field we perform a comparison between  the fluxes given by 
the GALEX pipeline, aperture photometry measurements and fluxes obtained by PSF fitting with DAOPHOT. 
From this comparison we conclude that the fluxes given by DAOPHOT and previously used might be slightly over-estimated by 0.2~mag. 
Although it is not clear if this shift is reliable for crowded objects we decided to apply the correction to all 
the UV data in the CDFS field: $\mbox{NUV(new)}= \mbox{NUV(old)} -0.2$~mag. 
This systematic correction remains very small (of the order of the error) and does not imply any 
modification of our previous results. 
Whereas the completeness of the NUV data at a
level of 80~\%  is obtained for $\mbox{NUV} = 26$~mag, we truncate the sample at $\mbox{NUV}=25.3$~mag in order that 
 more than 80~\% of the GALEX sources are identified in COMBO-17 with $R < 24$~mag. This limit ensures a redshift accuracy better than 10$\%$  \citep{wolf04}.
We also restrict the final sample
to objects with a single counterpart in COMBO-17 within 2 arcsec (i.e.\ 90~\% of the UV
sources).  300 galaxies are thus selected. 
The limit of $\mbox{NUV}=25.3$~mag corresponds to $\log L_{\rm UV} = 9.3 (L_{\odot})$ at $z=0.7$. 
44~\%  (131/300) of these sources are detected at 24 $\mu$m.  
For the undetected ones, we adopt an upper limit at 0.025~mJy  \citep{buat08}.  

At $z\simeq 1$  we must build a new sample. The NUV band of GALEX corresponds to 1155 \AA\ 
in the rest frame of the galaxies. 
Downward 1200~\AA\ the spectral energy distribution of galaxies is poorly known and the few available 
observations have  shown that the shape of the spectral energy distribution (SED) may vary 
a lot from galaxy to galaxy \citep{buat02, leitherer02}: the power-law model valid for 
$\lambda>1200$~\AA\ cannot be safely used  at shorter wavelengths. 
{}To avoid this difficulty we  perform a $U$ selection. 
We again work in the CDFS/GOODS field and  $U$-band observations have been performed as part 
of the EIS survey \citep{arn01}. 
We have cross-correlated the  $U$ selected catalogue with the COMBO-17 sample with a tolerance radius of 1 arcsec.  Redshifts are from COMBO-17+4 (Tapken, private communication) that provide safe redshifts, especially at $z \ge1$ in the GOODS-South field by combining COMBO-17 filters with three near infrared bands in ISAAC JHK bands.
69~\% of the $U$ sources are identified uniquely down to $U = 26$~mag. 
We cut the sample at $U = 24.3$~mag which corresponds to 80~\% of the sources identified in 
COMBO-17 with a single counterpart.  
The photometric redshifts from COMBO-17 are secure only for galaxies with $R<24$~mag (in the Johnson system). 
96~\% of our galaxies fill this condition.
We then select galaxies in the redshift bin 0.8--1.2 which results in a sample of 316 galaxies, at this redshift  
the $U$-band corresponds to $\sim 1800$~\AA\ in the galaxy rest-frame ($z=1$) when our reference 
wavelength is 1530~\AA. 
Because of the uncertainties about the shape of the UV SEDs we prefer to avoid any interpolation 
between the NUV and  $U$ observed fluxes and keep the uncorrected $U$ data. 
We can estimate the uncertainty due to the shift of the rest-frame wavelengths from 1800 to 1530~\AA\ 
by assuming a power-law for the continuum between these two wavelength, for reasonable values of 
the power-law valid for a UV selection ($ \alpha=-2~ \rm to~ 1$ with 
$F_{\alpha} \propto \lambda^{\alpha}$) the error is at most 20~\%. 
The limiting magnitude $U = 24.3$~mag corresponds to $\log L_{\rm UV}=9.9 ~(L_{\odot})$ at $z = 1$.
As for the sample at $z = 0.6\mbox{--}0.8$, the 24~$\mu$m data from SPITZER/MIPS  were 
cross-correlated to the $U$ sources identified in COMBO-17 within a tolerance radius of 2 arcsec \citep{buat08}. 
207 out of the 316 galaxies are detected at 24~$\mu$m. 
For the undetected ones, we again adopt an upper limit at 0.025~mJy  \citep{buat08}.

The aim of this paper is to determine if a UV selection is able to track all the star formation. The comparison of the IR and UV emissions will allow us to study star formation and dust attenuation affecting newly formed stars. In this context quasars and active galaxies are  excluded from all our samples. We exclude objects classified as QSO/Seyfert1  in the Combo-17 classification (2  objects at $z\simeq 0.7$ and 4 at $z\simeq 1$). We have also cross-correlated our galaxy sample with the X ray sources observed by CHANDRA in the field  \citep{bauer04} and  we discard 22 sources  at $z\simeq 0.7$ and 19 at $z\simeq 1$ as X-ray emitters. As a final check we can  compare mid-IR IRAC colours as suggested by \citet{stern05} to determine  if our samples were still  contaminated by AGNs. 1 and 4$\%$ of the samples at z=0.7 and 1 respectively were found in the AGN area close to the boundary where the contamination by star forming sources is significant. We choose not to exclude these remaining sources.

\subsection{Lyman Break Galaxies at $z\simeq 1$}

This paper presents a sample of galaxies in  the CDFS-GOODS field selected as being LBGs with 
a similar UV selection as that used in previous works (\citet{burgarella06}, \citet{burgarella07} and 
\citet{burgarella09}). 
The primary selection was performed in GALEX NUV band down to $\mbox{NUV}=25.3$~mag, 
as discussed in Section 2.1 this limit ensures us that 80~\% of the sources are identified with 
a reliable redshift in COMBO-17. 
Then  we look up into the GALEX FUV band for counterparts down to $\mbox{FUV}= 26.8$~mag 
(80~\% completeness on the GALEX detections) . To be selected as a Lyman Break Galaxy, 
an object must comply with the two criteria: 1) its redshift  must be in the range $0.9 < z < 1.3$ and 
2) its UV colour $\mbox{FUV}-\mbox{NUV} > 2$~mag. 
Indeed, some NUV faint objects that would be classified as LBGs if the FUV limiting magnitudes objects had 
reached down to $\mbox{FUV} < 27.3$~mag are not selected as LBGs in the present sample. 

This selection provides 117 LBGs with a unique counterpart in the optical out of which 58 are detected at 
24~$\mu$m, for the other ones an upper limit of 0.025~mJy is adopted as for the other samples. 
In addition, we have 33 LBGs with two counterparts in the optical out of which 20 are detected at $24\;\mu$m.

As for the $z=1$ sample, the UV luminosity is based on the observed $U$-band. 
At $z=1$ the selection of galaxies on their Lyman break focuses on galaxies with a high intrinsic UV 
continuum and a large intrinsic break since the role of the intergalactic medium on the amplitude 
of the break is known to be low at this redshift \citep[e.g.][]{malkan03}. The situation is different for LBGs selected at higher z whose break is dominated by the effect of the intergalactic medium. Nevertheless \citet{burgarella07} found similar spectral energy disributions for  LBGs at $z\simeq 1$ and $z\simeq 3$.

\subsection{ Estimating the IR  luminosities}

\begin{figure}
\centering
 \includegraphics[width=10cm]{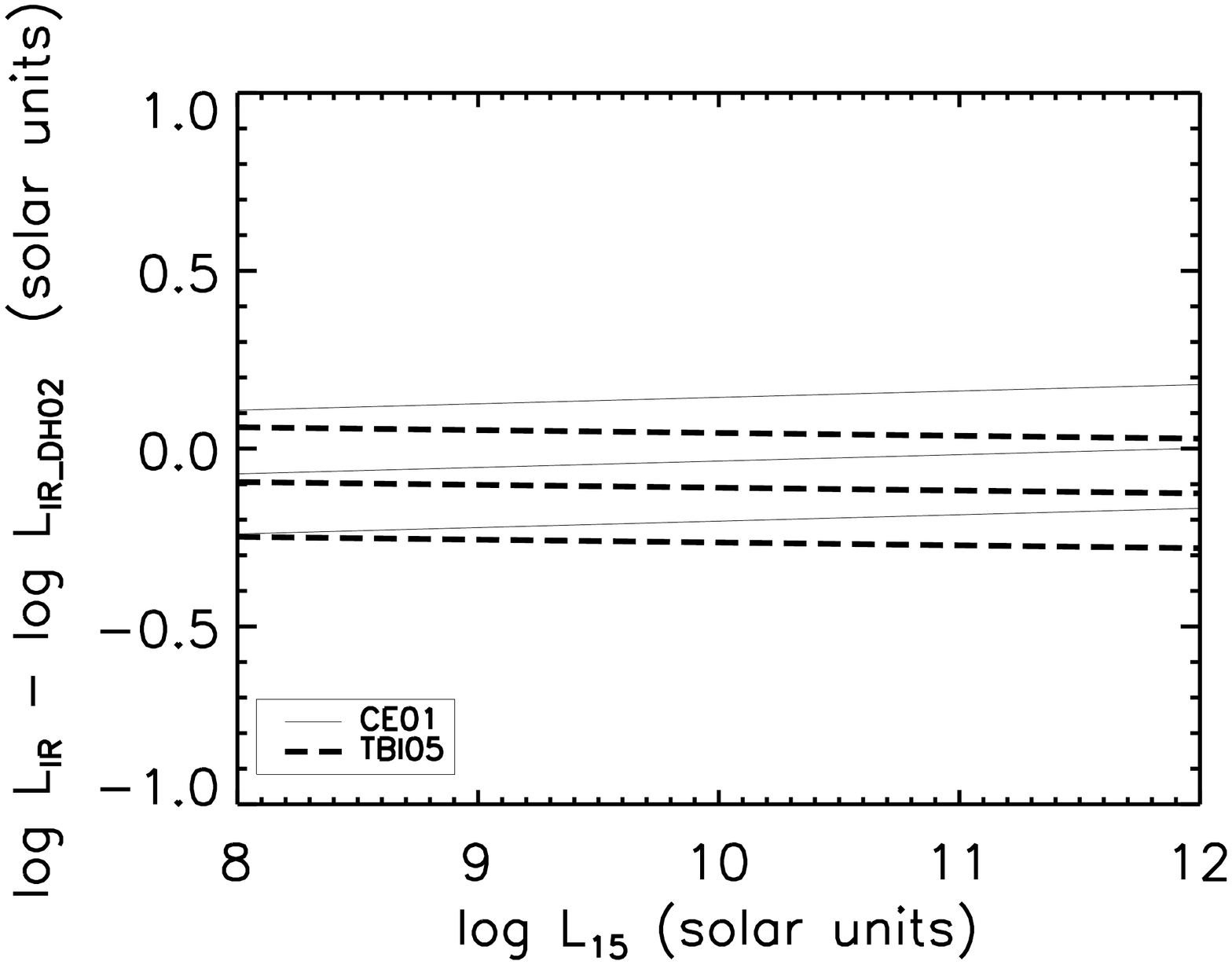}
 \includegraphics[width=10cm]{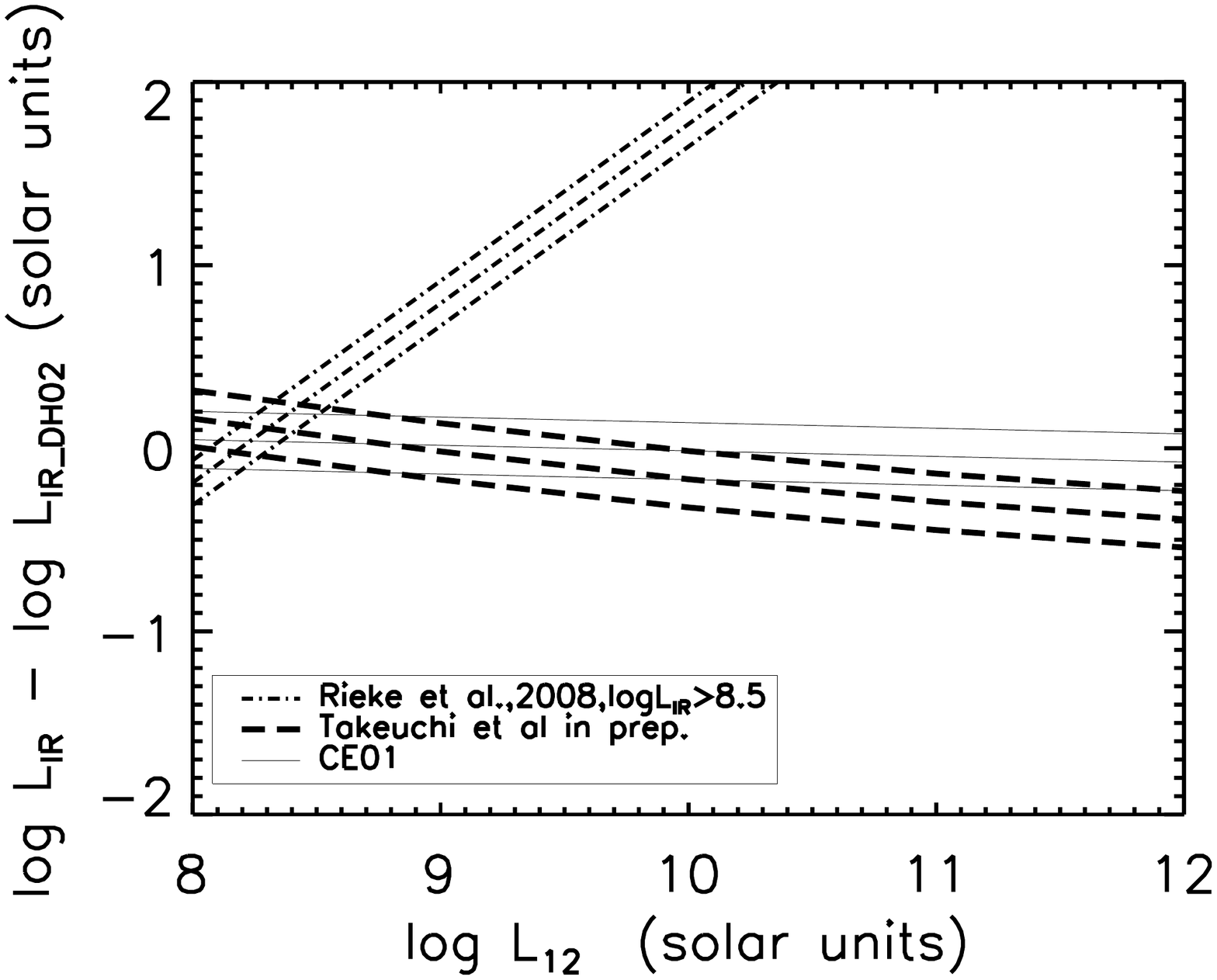}
  \caption{ Different calibrations of $ \log (L_{\rm 24})$ at $z = 0.7$  (upper panel)- and $z = 1$ (lower panel) versus $ \log L_{\rm IR}$. All the calibrations are normalized to the values found with the DH02 relations given in the text. The three lines plotted for each calibration correspond to the mean relation and the rms dispersion.}
      \label{calib-lir}
\end{figure}

The aim of the paper is to compare the IR and UV emissions of the galaxies from  $z=0$  to  $z \simeq 1$.
At $z=0$, the IR (8--1000~$\mu$m) emission is estimated using the calibration of \citet{dale} based on 
the 60 and 100 $\mu$m fluxes from IRAS \citep{buat05}. 
At higher $z$, we must  use the  emission at $24\;\mu$m to estimate the total far infrared emission. 
At $z=0.7$ (resp 1) the observed  $24\;\mu$m corresponds to  rest-frame $\simeq 15 \;\mu$m 
(resp $12 \;\mu$m). 
The extrapolation from the mid infrared (MIR)  emission to the total IR one is known to be quite difficult 
but up to $z=1$ we still remain in a wavelength range observed in the nearby universe by either IRAS or ISO. 
As a consequence several calibrations and IR SED templates were proposed in the literature 
based on observations of nearby galaxies by IRAS and/or ISO \citep[e.g.][]{chary01,dale02,ttt}.  
The recent observations of SPITZER have stressed the large variety of IR spectral energy distributions in nearby 
galaxies \citep{dale05,rieke08}, these observations   imply  large uncertainties in the extrapolation from a 
monochromatic flux to the total IR emission. 
{}To illustrate this uncertainty,  we  can compare  several calibrations of the monochromatic luminosities at 
12 and 15~$\mu$m into bolometric IR luminosities. 
We consider the calibrations of \citet{chary01} and \citet{dale02} based on templates built by combining IRAS 
and ISO data on small samples of galaxies. 
The \citet{dale02} templates are calibrated in total IR luminosities following the method of 
\citet{marcillac06}) and we obtain the following relations  between total and monochromatic luminosities:
\begin{equation}\label{eq:mir_tir12-1}
  \log L_{\rm IR} =  1.25  \log L_{\rm 12}-0.341 \;,
\end{equation}
\begin{equation}\label{eq:mir_tir12}
  \log L_{\rm IR} = 0.985 \log L_{\rm 15} +1.26\;
\end{equation}
We also consider the calibrations  of \citet{ttt} (hereafter TBI05) based on a statistical analysis of all 
the galaxies observed in the four bands of IRAS in the Point Source Catalog.  
The relation obtained by TBI05 at $\lambda = 12\; \mu$m for local galaxies
(i.e.\ when we observe a galaxy at $\lambda_{\rm obs}$, its emitted 
wavelength $\lambda_{\rm em}$ corresponds to $12\; \mu$m at $z = 1$) 
is slightly modified to take into account a slight non-linearity for the most luminous objects: 
\begin{equation}\label{eq:mir_tir12}
  \log L_{\rm IR} = 2.265+0.707 \log L_{\rm 12}+0.014 (\log L_{\rm 12})^2 \;,
\end{equation}
whereas at $\lambda = 15\;\mu$m, we used the original relation
\begin{equation}\label{eq:mir_tir15}
  \log L_{\rm IR} = 1.23+0.972 \log L_{\rm 15} 
\end{equation}
at $z = 0.7$, since the non linearity is not significant.
We have checked that the estimated $L_{\rm IR}$s do not differ by more 
than 5\% as compared to the old calibration presented in TBI05. 
 We also add to the comparison the recent calibration obtained by \citet{rieke08} at 12~$\mu$m and based on a compilation of SPITZER data with a special emphasis  on LIRGs and ULIRGs. The comparison between all these calibrations is shown in Fig.~\ref{calib-lir}.  In order to highlight the differences we have normalized the relations to those of DH02. The rms dispersion is overplotted for each relation.\\
At $15~\mu$m  (corresponding to the sample at $z = 0.7$ in the present work) the calibrations 
of  \citet{chary01} (hereafter CE01) and \citet{dale02} (hereafter DH02) are found very similar: 
the calibration of TBI05 leads to  a slightly lower $L_{\rm IR}$ by about 0.1 dex  than those 
obtained with the CE01 relation. 
This difference is much smaller than the intrinsic dispersion of the correlations and \citet{buat07-2} have 
found that both calibrations led to very similar results about dust attenuation for a sample of LIRGs. 
However, at $12\;\mu$m ($z=1$) the discrepancy is found  larger between the calibrations of 
CE01 and DH02 on the  one hand (again very similar) and that of TBI05 on the other hand: 
it reaches 0.2~dex for galaxies with $L_{\rm IR} \simeq 10^{11} L_{\odot}$ the TBI05 relation giving 
systematically lower IR luminosities.Nevertheless these relations  calibrated at $z=0$ remain marginally consistent within one rms 
 and we  expect at least the same amount of dispersion 
at higher redshift. 
The relation proposed by \citet{rieke08} for $ \log L_{\rm IR} >8.5 (L\odot)$  appears much  steeper than the other ones leading to larger IR luminosities 
especially for intrinsically luminous objects. 
\citet{rieke08} have gathered SEDs of  nearby LIRGs and ULIRGs together with the \citet{dale02} 
templates applied to the SINGS sample for galaxies of intermediate luminosity. 
Therefore we can infer that the discrepancy found between the \citet{dale02} and \citet{rieke08} calibrations 
is due to the introduction of these LIRGs and ULIRGs. 
Once again these differences illustrate the uncertainty inherent to these calibrations. 
At $z \simeq 0.7$, 26$\%$ of our galaxies detected  at 24~$ \mu$m  and 11$\%$ of the whole sample are LIRGs-ULIRGs; at $z\simeq 1$ these fractions reach 34$\%$ of the galaxies detected  at 24~$ \mu$m and 22 $\%$ of the whole sample. Although these fractions are significant, LIRGs-ULIRGs do not dominate our sample and we will not use the relation of 
\citet{rieke08}.  
We perform all the analyses reported in this work for the two calibrations TBI05 and CE01. 
The plots  are qualitatively similar and they will be presented with the TBI05 calibration. 
When quantitative evaluations are made (regressions or percentages) they will be given for both calibrations.

\section{Variation of $L_{\rm IR}/L_{\rm UV}$}

As already underlined in the introduction $L_{\rm IR}/L_{\rm UV}$ is a robust indicator of dust attenuation 
as long as we are dealing with galaxies forming stars actively. 
\citet{reddy08} and \citet{burgarella09} have reported a clear decrease of this ratio for LBGs at  $z=1$ 
and BM/BX galaxies at $z=2$ for a constant $L_{\rm IR}+L_{\rm UV}$ luminosity. Such a decrease of 
dust attenuation as redshift increases may have large consequences on the search of high redshift galaxies 
and the measure of their star formation rate. 
Here we re-investigate this question  with our homogeneous samples selected in a similar way at  $z=0$, 
$\simeq 0.7$ and $\simeq 1$.

\subsection{$L_{\rm IR}/L_{\rm UV}$ versus $L_{\rm IR}+L_{\rm UV}$}

\begin{figure}
\centering
\includegraphics[width=12cm]{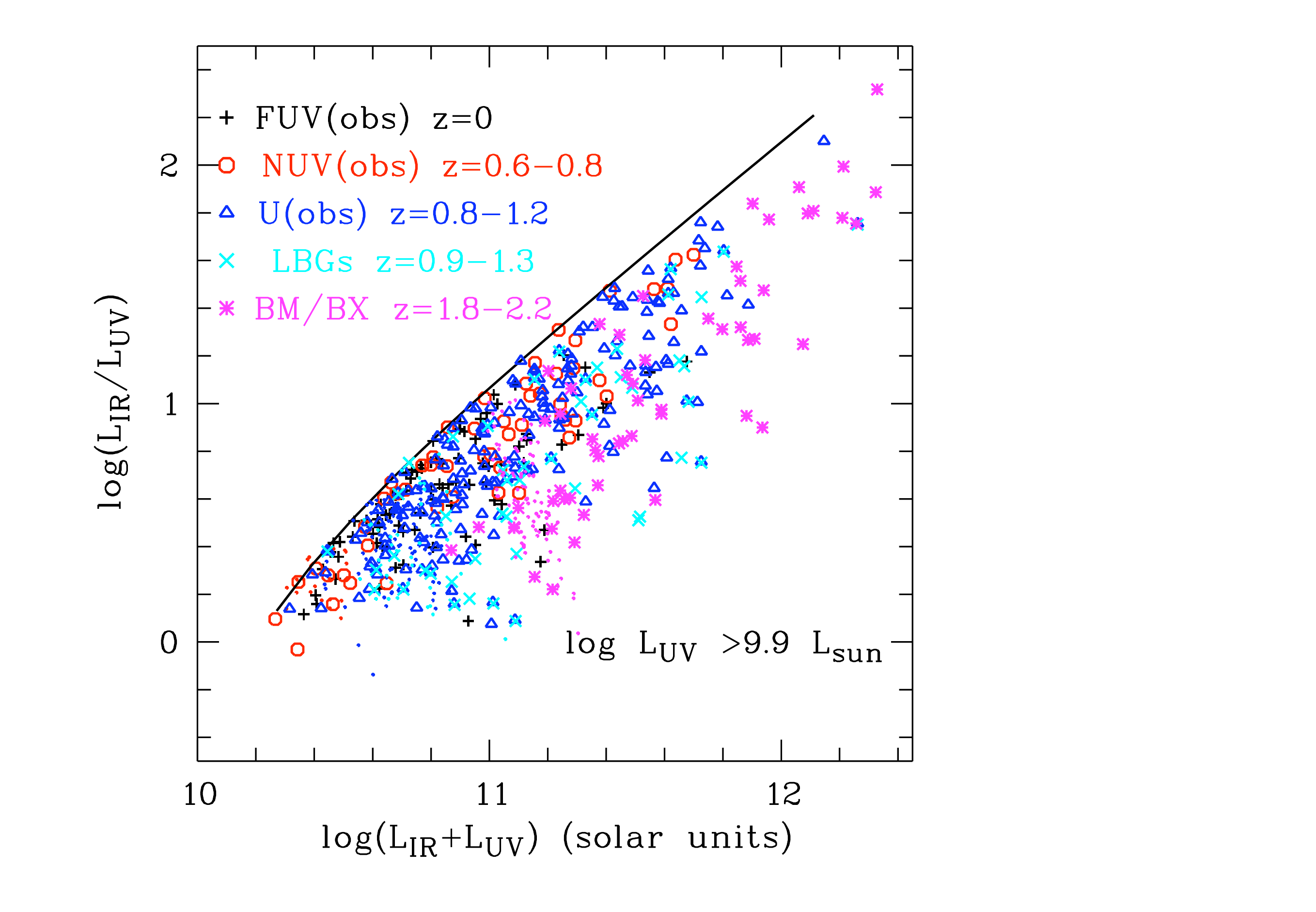}
  \caption{$ \log(L_{\rm IR}/L_{\rm UV})$ versus $ \log(L_{\rm IR}+L_{\rm UV})$ for the different samples (i.e. redshifts) defined in this work and the BM/BX galaxies of \citet{reddy06}. The different symbols corresponds to detections at 24~$\mu$m ($z=0$: plus, $z=0.6\mbox{--}0.8$: circles, $z=0.8\mbox{--}1.2$: triangles, LBGs: crosses, BM/BX: stars), dots are for  upper limits. The solid line corresponds to the adopted limit in UV luminosity: $ \log (L_{\rm UV} [L_{\odot}] = 9.9$).}
      \label{compextinc}
\end{figure}

\begin{figure}
\centering
 \includegraphics[width=12cm]{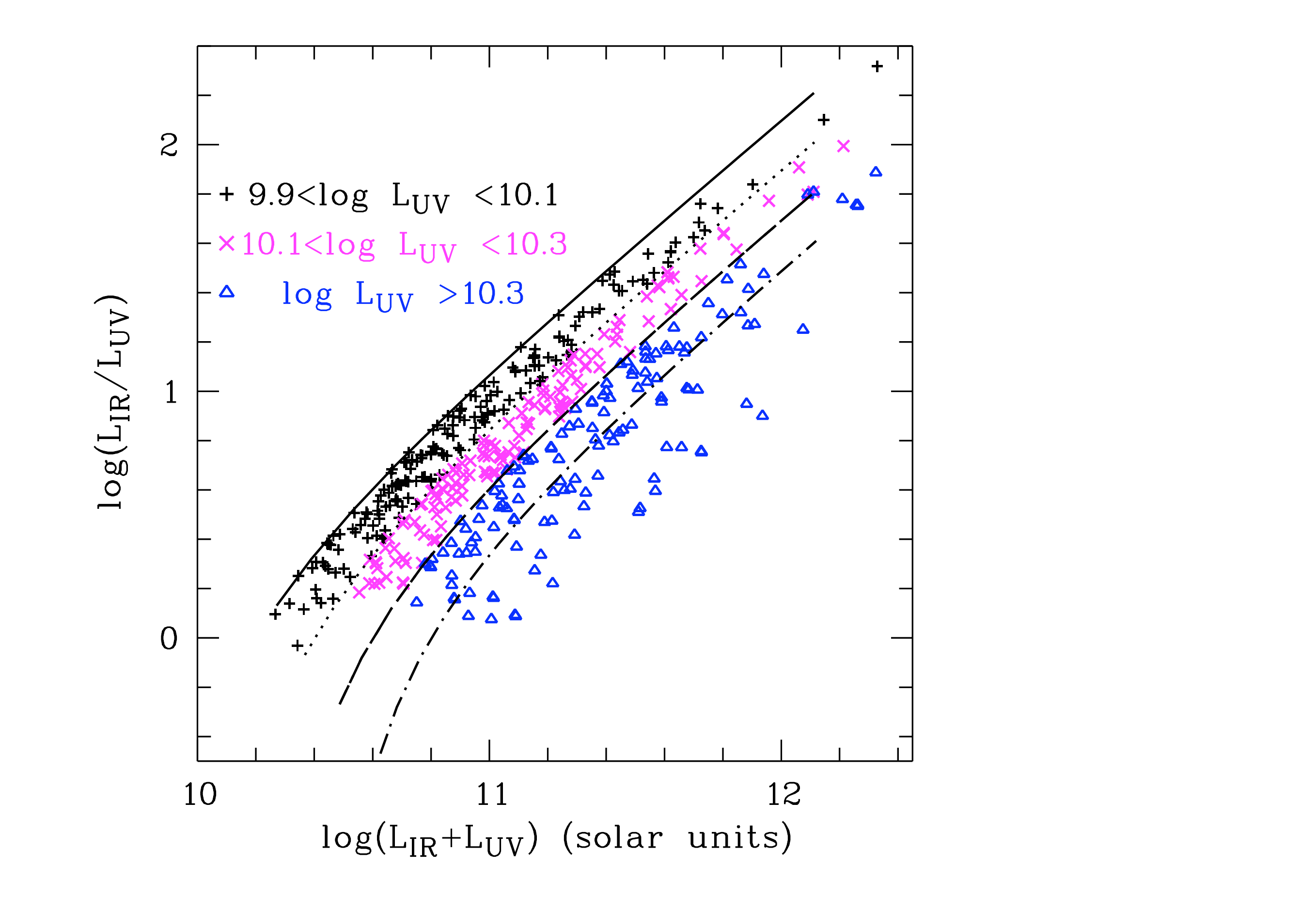}
  \caption{ $ \log(L_{\rm IR}/L_{\rm UV})$ versus $ \log(L_{\rm IR}+L_{\rm UV})$, 3 sub-samples are defined according to the UV (rest-frame) luminosity and different symbols are used for each of them. The lines correspond to the locus of galaxies with a fixed   $L_{\rm UV}$:  $ \log L_{\rm UV}=9.9$ (solid line), 10.1 (dotted line), 10.3 (dashed line) and 10.5 (dot-dashed line) in solar units.  For the sake of simplicity upper limits at 24 $\mu$m are not over-plotted }
      \label{extinclum}
   \end{figure}
\begin{figure}
\centering
 \includegraphics [width=12cm]{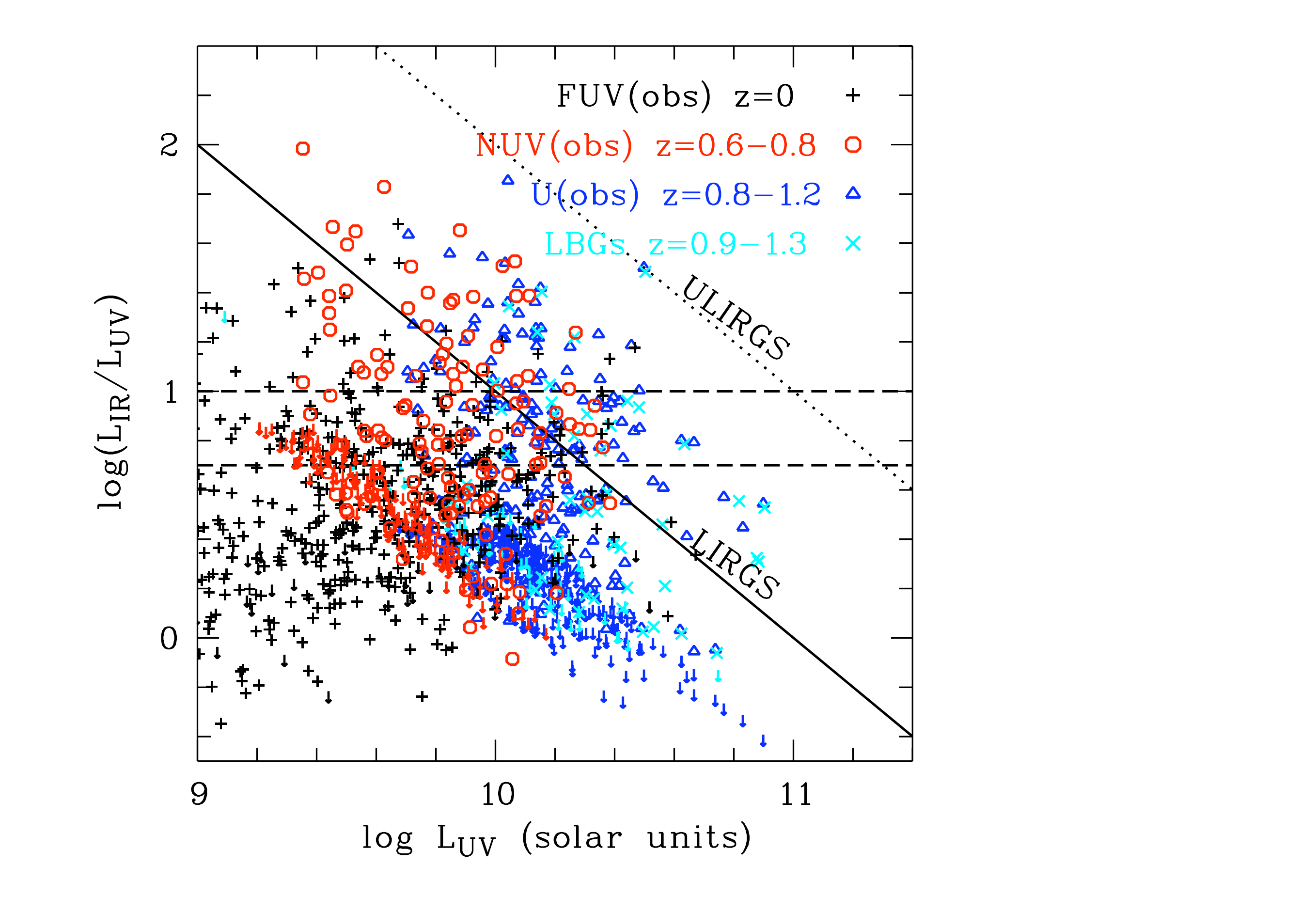}
   \caption{ $ \log(L_{\rm IR}/L_{\rm UV})$ versus $L_{\rm UV}$ for the different samples defined in this work. The diagonal solid lines are the limits above which galaxies are LIRGs or ULIRGs.}
      \label{plotluv}
\end{figure}

\begin{figure}
\centering
 \includegraphics [width=12cm]{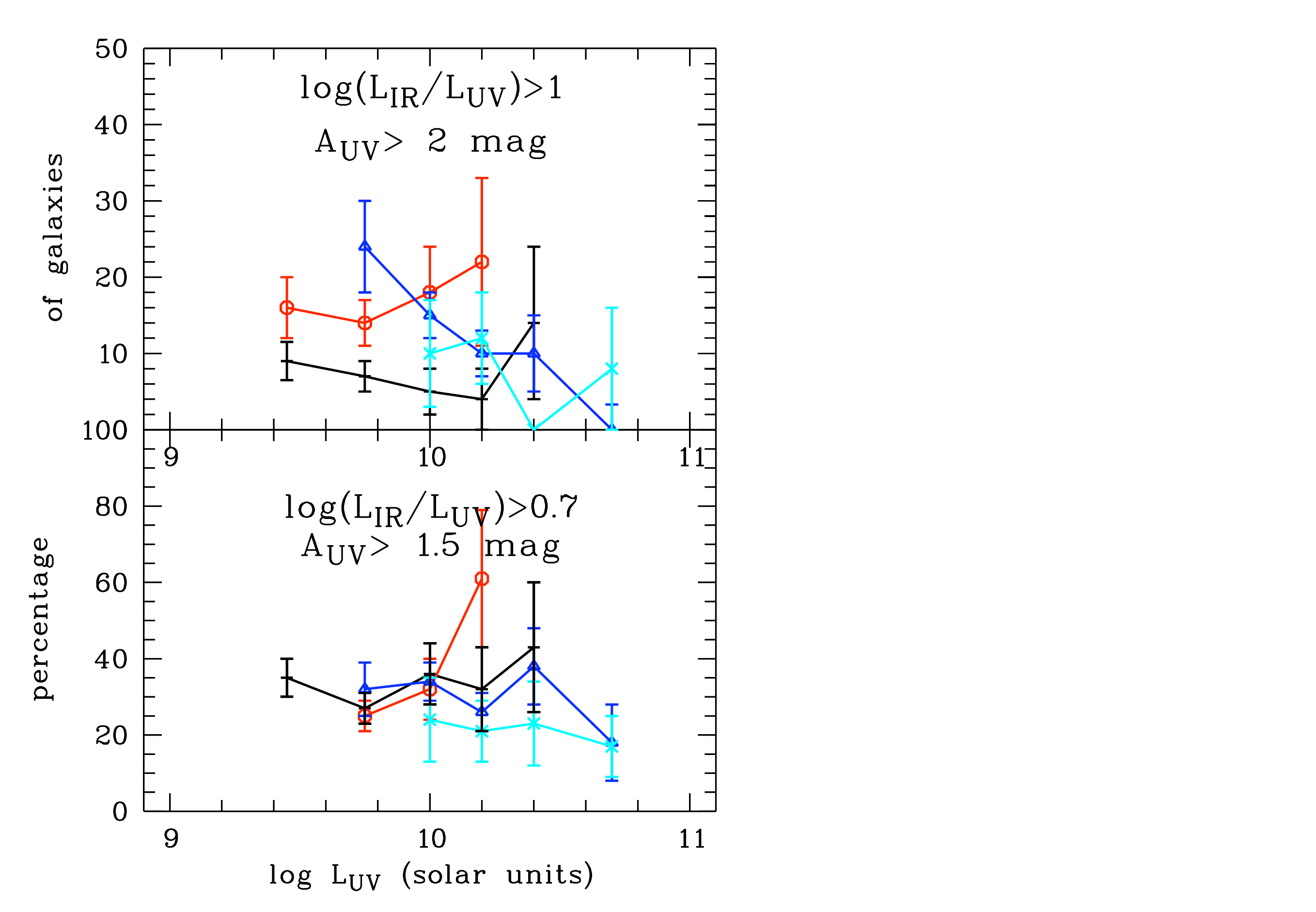}
  \caption{fraction of galaxies with $ \log(L_{\rm IR}/L_{\rm UV})>0.7$, $ \log(L_{\rm IR}/L_{\rm UV})>1$. The symbols are the same as in Fig.~\ref{compextinc}. The first bin of luminosity in the lower panel   is not considered at $z=0.7$  since the upper limits of $\log(L_{\rm IR}/L_{\rm UV})$ calculated for the galaxies with this luminosity and undetected at 24~$\mu$m may be higher than 0.7. }
      \label{pourcents}
   \end{figure}
\begin{figure}
\centering
\includegraphics [width=10cm]{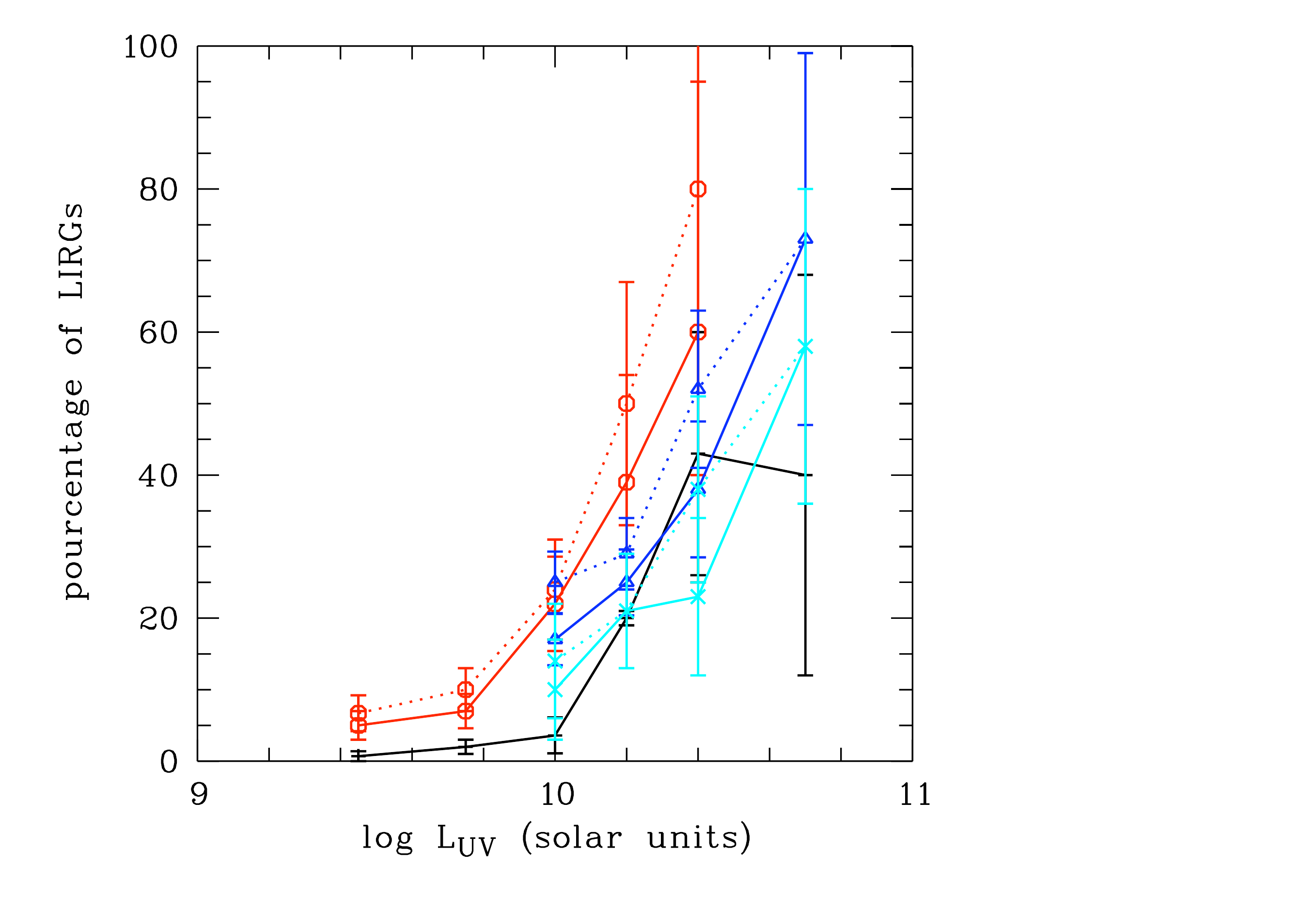}
   \caption{ fraction of LIRGs, the symbols  are the same as in Fig.~\ref{compextinc}. The solid lines refer to the TBI05 calibration for $L_{\rm IR}$ and the dotted lines to the CE01 calibration}
      \label{statlirg}
\end{figure}

As detailed in the previous section the samples are all selected in UV rest-frame at wavelengths sufficiently 
close to avoid  $K$-corrections. 
Nevertheless, the range of luminosities observed in each sample is different.
Before any comparison we must cut the samples at the same luminosity. The most stringent limit is for the 
$U$-selected sample. With a cut at $U=24.3$~mag we are only able to observe galaxies with 
$\log L_{\rm UV} > 9.9 ~(L_{\odot})$ at $z=1$.
In Fig.~\ref{compextinc} are reported all the galaxy samples considered in this work and truncated at 
$\log L_{\rm UV}>9.9~ (L_{\odot})$. A trend with redshift seems to be present with galaxies shifted toward 
the right when  $z$ increases. The mean values found for  $L_{\rm IR}/L_{\rm UV}$  in the luminosity bin 10.8-11.3 $(L_{\odot})$ are  $<L_{\rm IR}/L_{\rm UV}>=0.86\pm 0.35, 1.07\pm 0.30,0.75\pm 0.30 ~{\rm and} ~0.54\pm 0.35$ at  $z=0,0.7,1$ and for the LBG sample respectively. 
Although no clear trend is confirmed from z=0 to 1 given the large dispersion of the $L_{\rm IR}/L_{\rm UV}$ distribution,   for a given $L_{\rm IR}+L_{\rm UV}$, dust attenuation as traced by $L_{\rm IR}/L_{\rm UV}$  
  is lower for   LBGs at $z\simeq 1$.\\
We have also over plotted the sample of   BM/BX galaxies of \citet{reddy06}  at z=2. The IR luminosities are  estimated from the 24 $\mu$m fluxes using the calibration of \citet{caputi07} as  in \citet{reddy08} and  the UV luminosity is calculated with the G band fluxes. Given the differences between these estimates and those performed for the samples studied in this work we do not make a quantitative comparison.  Anyway the BM/BX galaxies continue the trend reported  for LBGs at $ z\simeq 1$.

However the interpretation of this plot is made difficult by the fact that the quantities reported on the axes 
are both a combination of $L_{\rm IR}$ and  $L_{\rm UV}$. 
This is illustrated  in Fig.~\ref{extinclum} where  the galaxies are considered according to 
their UV luminosity, whatever their redshift is. 
The lines represent the locus of galaxies with a given $L_{\rm UV}$. 
The locus of the galaxies in this plot is strongly constrained by their UV luminosity: when 
$L_{\rm IR}/L_{\rm UV}$ varies, galaxies with a given $L_{\rm UV}$ move along  lines like  those 
over-plotted on the diagram.
Therefore the shift seen in Fig.~\ref{compextinc} is due to the presence of more luminous galaxies in UV when 
$z$  increases as expected from the evolution of the UV luminosity function \citep{arnouts05,ttt06}. 
The variation of dust attenuation as traced by $L_{\rm IR}/L_{\rm UV}$ can only be quantified along  
these lines of constant $L_{\rm UV}$.

\subsection{$L_{\rm IR}/L_{\rm UV}$ versus $L_{\rm UV}$}

{}From the above analysis, it is clear that we must avoid to combine  IR and UV luminosities on both 
axes since the resulting plot is  too strongly constrained. 
We can analyse the variation of  $L_{\rm IR}/L_{\rm UV}$ as a function of $L_{\rm UV}$ alone.  
In such a plot  we are not  affected by volume effects  since all the samples are selected in UV. 
In Fig.~\ref{plotluv} is reported the variation of $L_{\rm IR}/L_{\rm UV}$ as a function of $L_{\rm UV}$.
The loci of LIRGs ($L_{\rm IR}>10^{11}~L_{\odot}$) and ULIRGs ($L_{\rm IR}>10^{\rm 12}~L_{\odot}$) are 
also indicated. 
This time the general shape of this diagram is strongly constrained by the upper limits  at 24~$\mu$m 
which hamper any discussion about the low values of $L_{\rm IR}/L_{\rm UV}$. 
The upper envelope of the distribution shows a trend: as $L_{\rm UV}$ increases the maximum value of 
$L_{\rm IR}/L_{\rm UV}$  decreases from $\sim 50$ to $\sim 3$ ($\sim 65$ to $\sim 5$ for the CE01 calibration). 
A quantitative interpretation of this varying upper limit is difficult because of the statistics: 
since the total number of galaxies per luminosity bin decreases as the UV luminosity increases,  
we expect less extreme cases even with a similar parent distribution for $L_{\rm IR}/L_{\rm UV}$. 
Anyway the most UV luminous  galaxies exhibit a very moderate dust attenuation ($\log(L_{\rm IR}/L_{\rm UV}) = 0.5$ 
corresponds to $A_{\rm UV} = 1.2$~mag with the calibration of \citet{buat05}) and the galaxies with 
the largest dust attenuation are the faintest ones in UV. \\
In order to go further in the interpretation of Fig.~\ref{plotluv} we have calculated the fraction of galaxies with $ \log(L_{\rm IR}/L_{\rm UV})$ larger than 0.7 and 1 (corresponding to $A_{\rm UV} = 1.5$ and 2 mag respectively \citep{buat05}) as well as the fraction of LIRGs for each redshift, as a function of $L_{\rm UV}$. The cuts adopted for $\log(L_{\rm IR}/L_{\rm UV})$ (0.7 and 1) are chosen to be not affected by the non detections at 24  $\mu$m in the high redshift samples (cf the upper limits reported in Fig.~\ref{plotluv}) . If the  CE01 calibration is used  instead of the TBI05 one  the cuts  in $\log(L_{\rm IR}/L_{\rm UV})$ have to be increased of 0.1~dex (i.e., $\log(L_{\rm IR}/L_{\rm UV})=$0.8 and 1.1) . 

The results are reported in Fig.~\ref{pourcents} and ~\ref{statlirg}.  
At $z=0$ all the galaxies are detected in IR and  the distribution of $L_{\rm IR}/L_{\rm UV}$ is well described 
by a Gaussian with a mean value of 0.55 dex and a standard deviation of 0.3 dex.
If we first consider the fraction of galaxies with $\log(L_{\rm IR}/L_{\rm UV})>0.7$, this fraction is not found 
to be very dependent on the redshift or the UV luminosity at least up to 
$\log L_{\rm UV} \le 10.3\mbox{--}10.4~(L_{\odot})$ (there is only a "discrepant" point at $z=0.7$ but 
with a very large error bar). 
For the highest observed UV luminosities ($\log (L_{\rm UV} [L_{\odot}]) \ge 10.3\mbox{--}10.4$), 
only present in the samples at $z\simeq 1$, the fraction of galaxies with $\log(\rm L_{\rm IR}/L_{\rm UV}) > 0.7$  
decreases both for the  $U$ selection and the LBGs. 
The LBG sample has  a slightly lower fraction of  galaxies with $\log(L_{\rm IR}/L_{\rm UV})>0.7$ than that 
found in the  $U$ selected sample over  the whole range of luminosity. 

The fraction of galaxies with $\log(L_{\rm IR}/L_{\rm UV})>1$ highlights galaxies with the largest dust attenuation. 
This fraction does not exceed $\simeq 20\;\%$ for all our samples. Galaxies with such a high extinction 
seem to be more  frequent at $z > 0$ than at $z=0$: the distribution of  $L_{\rm IR}/L_{\rm UV}$ is  found  to 
reach larger values  at $z  > 0$ than at $z = 0$ but we must remain cautious because on the uncertainties 
on the MIR-total IR  calibration. 
As found above , at $z=1$  there is almost no UV luminous  galaxy with a large attenuation and the fraction 
of galaxies with $\log(L_{\rm IR}/L_{\rm UV})>1$ increases toward lower UV luminosities.

The evolution of the fraction of LIRGs  is reported in Fig.~\ref{statlirg}. 
This  fraction  increases with the UV luminosity: it is expected even  even without any evolution of the 
$L_{\rm IR}/L_{\rm UV}$  distribution. 
For the UV selected galaxies at  $z=0$, 0.7 and 1, the variations are found similar with a slightly larger 
fraction of LIRGs for a given UV luminosity at $z>0$, it is indeed the same effect as noted for 
galaxies with $\log(L_{\rm IR}/L_{\rm UV})>1$. 
The fraction of LIRGs in the LBG sample is systematically lower than that found for the UV selected 
galaxies at $z=1$, again leading to the conclusion of a lower dust attenuation for these galaxies.

 \citet{reddy08} estimated the colour excess distribution of  BX galaxies at $z\simeq 2$ and LBGs at $ z\simeq 3$. They found   $<E(B-V)>= 0.15\pm 0.07$. Adopting the dust attenuation law of \citet{calzetti00} gives   $<A_{\rm UV}> = 1.5$mag. As a consequence 50$\%$ of  BX galaxies and LBGs have  $A_{\rm UV} > 1.5$  mag and if we assume that the distribution of E(B-V) is Gaussian, 30$\%$ have  $A_{\rm UV} > 2$ mag. \citet{reddy08} obtained similar results by an analysis of   the $L_{\rm IR}/L_{\rm UV}$ distribution. They also found an average dust attenuation which does not vary with the UV rest-frame luminosity.

Therefore dust attenuation in UV selected galaxies at $z>1$ seems to be slightly larger than that found at $z\simeq1$, the difference being particularly significative for UV luminous galaxies for  there is   a hint for a lower dust attenuation at $z\simeq 1$. Nevertheless  we must remain cautious in our conclusions given the uncertainties in the  estimates of  dust attenuation and the different methods adopted: IR to UV flux ratio up to $z=1$  based on 12 and 15 $\mu$m luminosities, UV colours  and 8 $\mu$m luminosities at higher z.  These methods are known to give different results at least up to $z=1$ \citep{burgarella07, elbaz07} and the calibration of the  8 $\mu$m luminosity in total IR luminosity is highly uncertain \citep{caputi07, burgarella09}.  
\begin{figure*}
\centering
\includegraphics [width=15cm]{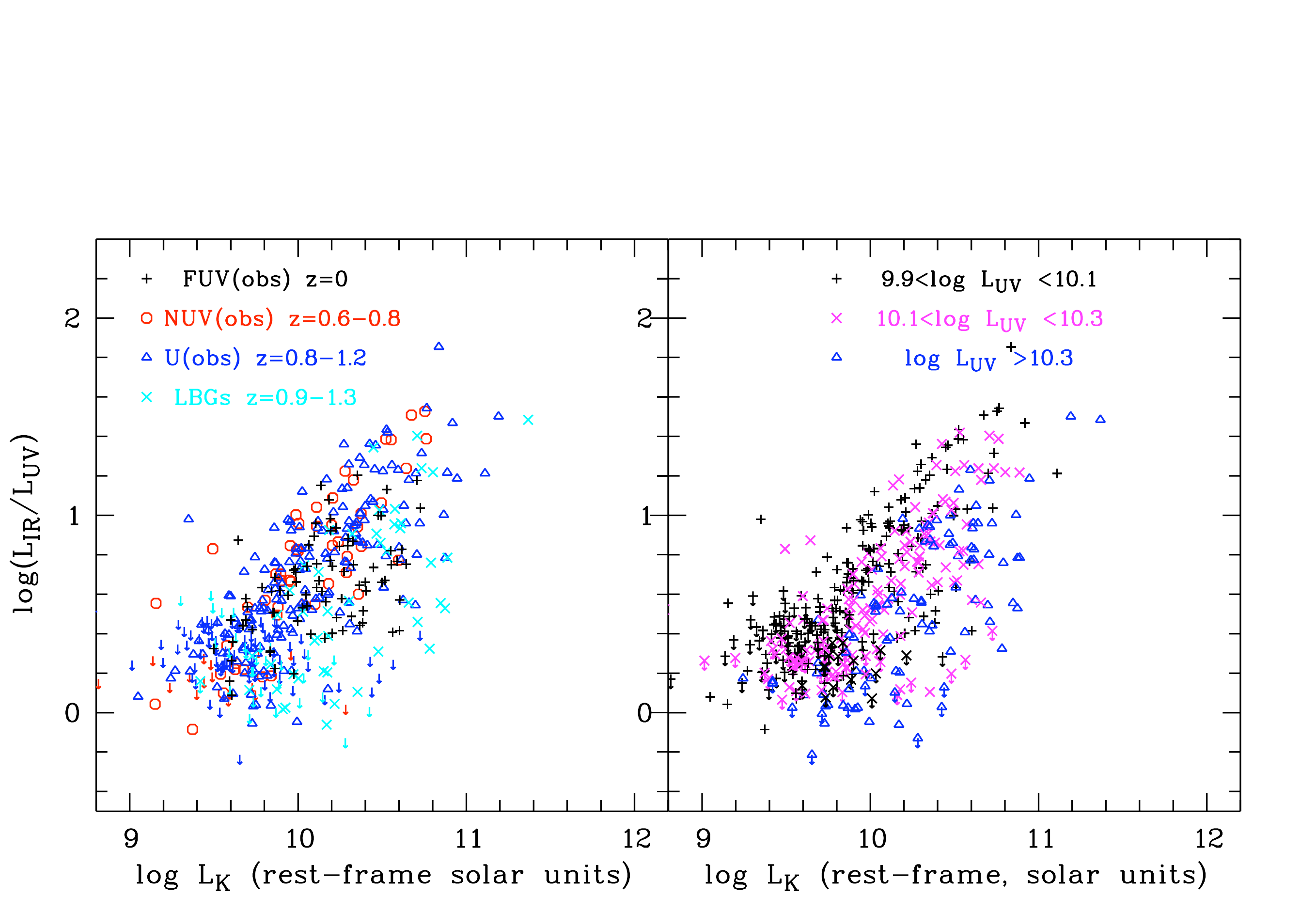}
   \caption{ $L_{\rm IR}/L_{\rm UV}$ versus the $K$ rest-frame luminosity $L_K$ expressed in solar units. Left panel: the different samples are plotted with the same symbols as in Fig.~\ref{compextinc}. Right panel: the samples are split according to the UV luminosity as in Fig.~\ref{extinclum}}
      \label{plotlk} 
\end{figure*}

\subsection{$L_{\rm IR}/L_{\rm UV}$ versus $L_K$}

We have seen that it is difficult to interpret the variation of $L_{\rm IR}/L_{\rm UV}$ as a function of a quantity also depending on these two luminosities. We can also use a quantity independent of them to avoid such an issue. In this section we will consider the rest-frame $K$ luminosity of the galaxies which is a tracer of the stellar mass of galaxies.  They are calculated with the IRAC band at 3.6 and 4.5~$\mu$m which corresponds to rest-frame $K$ at $z=0.7$ and 1  and with 2MASS data at  $z=0$  \citep{buat07-1}
In Fig.~\ref{plotlk} is reported the variation of $L_{\rm IR}/L_{\rm UV}$ as a function of  $L_K$ for the different samples and also split in luminosity bins. A net increase of $L_{\rm IR}/L_{\rm UV}$ with $L_K$ is found without a clear evolution with  $z$  for the galaxies selected in UV rest-frame,  only LBGs appear to have a lower dust attenuation for a given  $L_K$ .
When the samples  are split according to the UV luminosity of the galaxies it appears in average that  the more UV luminous  objects   exhibit a lower $L_{\rm IR}/L_{\rm UV}$ for a given $K$ luminosity than UV fainter sources. This is in agreement with what has been found in section 3.2. A linear regression on both $L_{\rm UV}$ and $L_K$ gives 
\begin{eqnarray}
  &&\log(L_{\rm IR}/L_{\rm UV}) = \nonumber \\
  &&\quad 0.78(0.04)\log L_K-0.79 (0.06) \log L_{\rm UV}+0.86(0.21)
\end{eqnarray}
Excluding LBGs from the analysis (since they are less extinguished  than UV selected galaxies) leads to a slightly different regression
\begin{eqnarray}
  &&\log(L_{\rm IR}/L_{\rm UV}) = \nonumber \\
  &&\quad 0.78(0.04)~\log L_K-0.65(0.07) \log L_{\rm UV}-0.58(0.21)
\end{eqnarray}
Using the calibration of \citet{chary01} would give  
\begin{eqnarray}
  &&\log(L_{\rm IR}/L_{\rm UV}) = \nonumber \\
  &&\quad 0.81(0.04)~\log L_K-0.54 (0.06) \log L_{\rm UV}-1.92 (0.26)
\end{eqnarray} for the whole sample and 
\begin{eqnarray}
  &&\log(L_{\rm IR}/L_{\rm UV}) = \nonumber \\
  &&\quad 0.81(0.04)~\log L_K-0.54 (0.08) \log L_{\rm UV}-1.91 (0.25)
\end{eqnarray} 
when LBGs are excluded.

\citet{martin07} and \citet{iglesias07} also studied the variation of $L_{\rm IR}/L_{\rm UV}$ as a function of the stellar mass from $z = 0$ to 1 for UV selected galaxies, our results are globally consistent with theirs: the relation between $L_{\rm IR}/L_{\rm UV}$ and $L_K$ or $ M_{\rm star}$ exhibit the same steepness, the shift with redshift reported  in these studies   is of similar amplitude to the one we find as a function of $L_{\rm UV}$. The main differences between these previous analyses and ours is that we select galaxies with the same limit in UV luminosity whereas the other studies were based on magnitude limited samples with a different limit in luminosity when  $z$  varies. Our approach allows us to emphasise differences of behaviour as a function of the luminosity of the galaxies and this effect seems to be at the origin of the redshift evolution reported earlier.

 \begin{figure*}
\centering
\includegraphics [width=15cm]{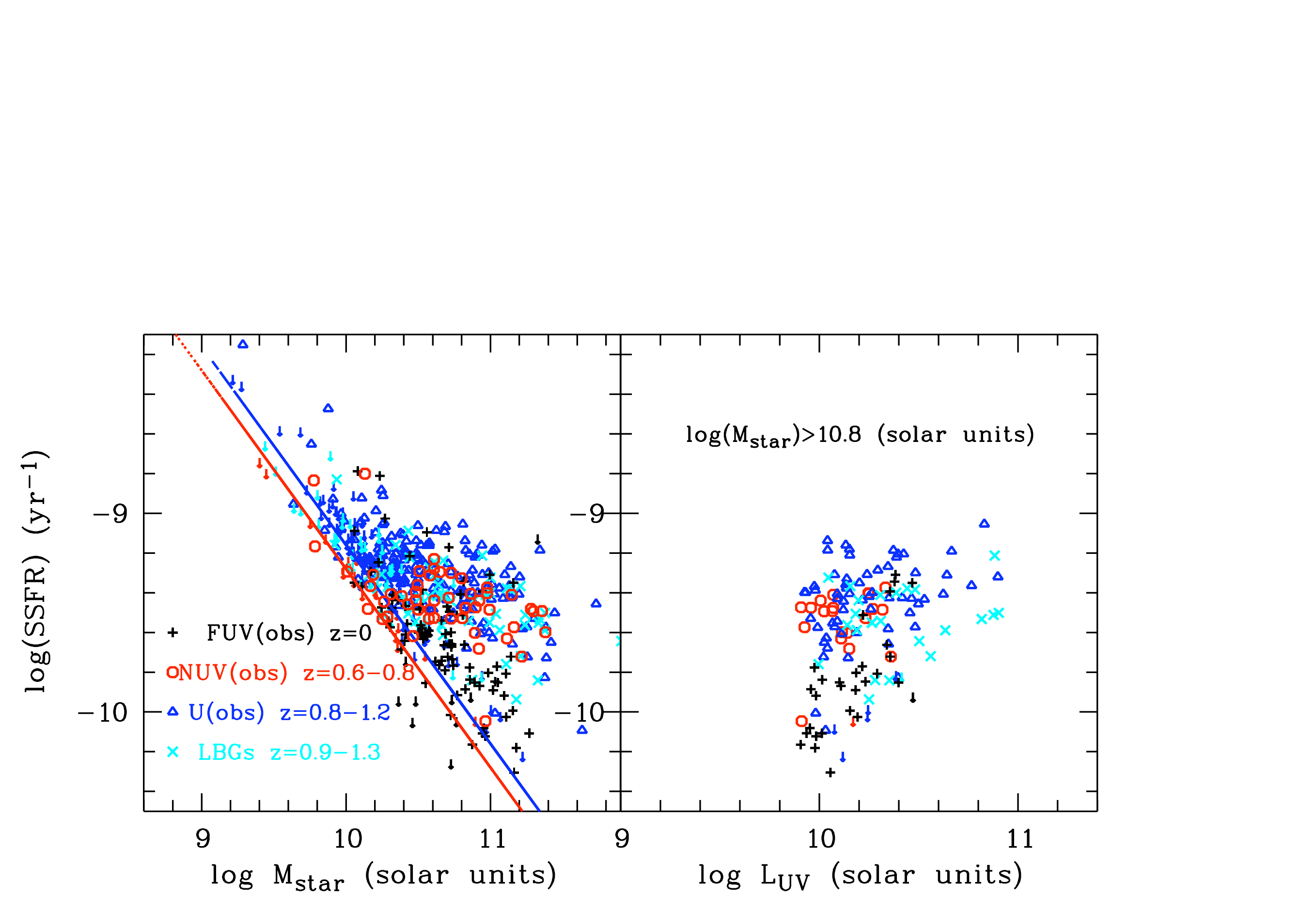}
   \caption{ Variations of the specific star formation rate for every sample considered in  this work, all the samples are truncated at $\log L_{\rm UV} > 9.9$ in solar units. Left panel:  as a function of the stellar mass of the galaxies . The diagonal lines are the lower limits in SSFR for $z = 0.7$ (lower line) and 1 (upper line). Right panel: as a function of the UV luminosity for galaxies  with $\log( M_{\rm star})> 10.8 (M_{\odot})$ }
      \label{ssfr}
   \end{figure*}
   
\section {Star formation activity}

Dust attenuation  has been found to be related to the observed UV and $K$ luminosity of our 
UV selected galaxies, LBGs exhibiting more extreme behaviour. Is the strength of the star formation activity also 
linked to the UV luminosity? 
Do LBGs exhibit different star formation activity than UV selected galaxies?
The star formation activity can be quantified by the specific star formation rate defined as the current 
star formation rate divided by the stellar mass of the galaxies. We can estimate this quantity with our data sets.  
We calculate the total SFR   by adding the SFR from the IR and the UV (observed) luminosities 
\citep{iglesias06,buat08}. We adopt a Salpeter IMF and the formulae of \citet{iglesias06}:
\begin{eqnarray}
  &&\log {\rm SFR}_{\rm IR} (M_{\odot} {\rm yr^{-1}}) = \log L_{\rm IR}(L_{\odot})-9.75 \\
  &&\log {\rm SFR}_{\rm UV}  (M_{\odot} {\rm yr^{-1}}) = \log L_{\rm UV}(L_{\odot})-9.51 \;.
\end{eqnarray}

The total SFR is expressed as $ \rm SFR_{\rm IR}+SFR_{\rm UV}$ except at  $z=0$ where the contribution 
of the dust emission not related to the star formation is estimated to be 30~\% \citep{iglesias06}.
The stellar masses of the galaxies are calculated with the IRAC band at 3.6 and $4.5~\mu$m which 
corresponds to rest-frame $K$-band at $z = 0.7$ and 1 and with the calibration of \citet{arnouts07}. 
At $z=0$, the calibration of \citet{bell03} is used  as discussed in \citet{iglesias06} and \citet{buat08}. 
We adopt a Salpeter IMF and check that the extrapolation of the calibration of \citet{arnouts07} is consistent 
with that we adopt at $z=0$ within 30~\% (0.1~dex). 

In Fig.~\ref{ssfr} is reported the variation of the SSFR as a function of the stellar mass for each sample 
truncated at $\log L_{\rm UV} > 9.9 ~(L_\odot)$. 
Such a limit combined with the detection limit adopted for the fluxes at $24\;\mu$m (25 $\mu$Jy) gives 
a limit in SFR for $z=0$ and 1 also indicated in Fig.~\ref{ssfr}. 
The SSFR at a given stellar mass  increases with $z$ as reported in both observational and theoretical 
studies and predicted  in scenarios of galaxy evolution.
The SSFR also exhibits a flat distribution: it is expected when only star forming galaxies 
are selected \citep{elbaz07, buat08}. 
The consistency between models and the mean  trends found in  UV and IR selected samples has been 
shown to be good up to $z=0.7$ \citep{buat08} but breaks at $z\ge 1$ \citep{elbaz07}. The purpose of 
this paper is not to perform a comparison between models and observations since we are dealing with 
only a sub sample of the overall galaxy population: the objects with $\log L_{\rm UV} > 9.9 (L_{\odot})$. 
We want to  compare the properties of these galaxies at different  $z$  and with LBGs.  
LBGs and UV selected galaxies seem to experience similar SSFRs at the same redshift.  
 We now focus on the most massive galaxies with  $\log( M_{\rm star})> 10.8 (M_{\odot})$ for which the detection limits reported in  Fig.~\ref{ssfr} do not induce a substantial bias.  The  galaxies of this sub-sample with a moderate UV luminosity ($\log L_{\rm UV} <10.4 (L_\odot)$) exhibit a large range of SSFRs, the most quiescent objects being found at z=0. Conversely all the most UV luminous objects ($\log L_{\rm UV} >10.4 (L_\odot)$)  exhibit quite large SSFRs. 
These galaxies  experiment  high SFRs  between 10 and 190  $M_{\odot} \rm yr^{-1}$ with an average 
	value of $55~M_{\odot} \rm yr^{-1}$: with such a rate they might have formed all their mass in few Gyr. 
\section{Total UV+IR luminosity functions }

Do we miss star forming galaxies in a UV selection up to $z = 1$, and as 
a consequence are we able to retrieve all the star formation when 
applying a reliable dust attenuation to galaxies selected in UV rest-frame? 
{}To answer these questions, it would be the most direct to construct 
the luminosity functions (LFs) with total luminosity related to star 
formation activity, $L_{\rm UV} + L_{\rm IR}$.

\subsection{Method}

\begin{figure}[h]
\centering
\includegraphics[width=10cm]{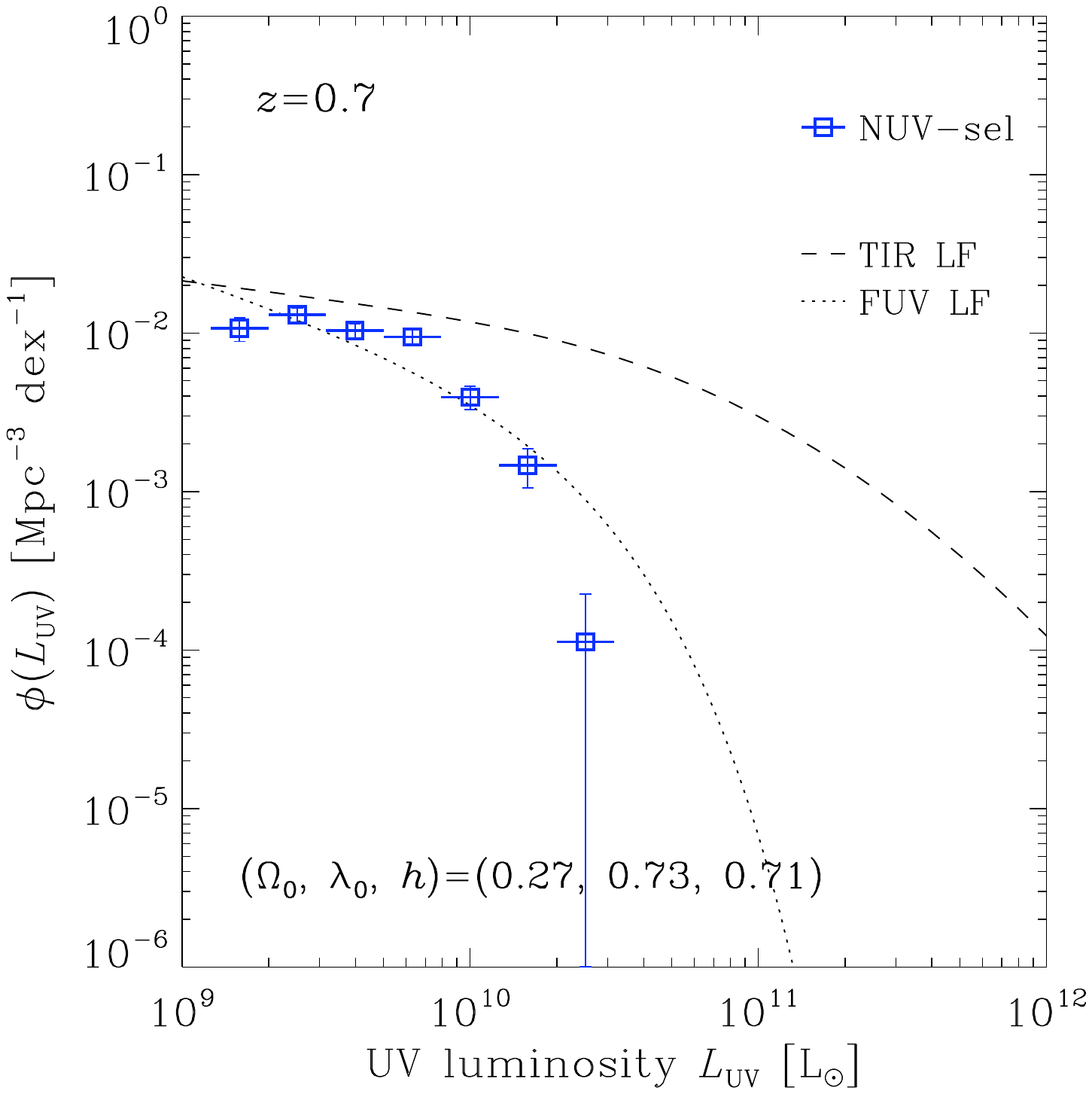}
\includegraphics[width=10cm]{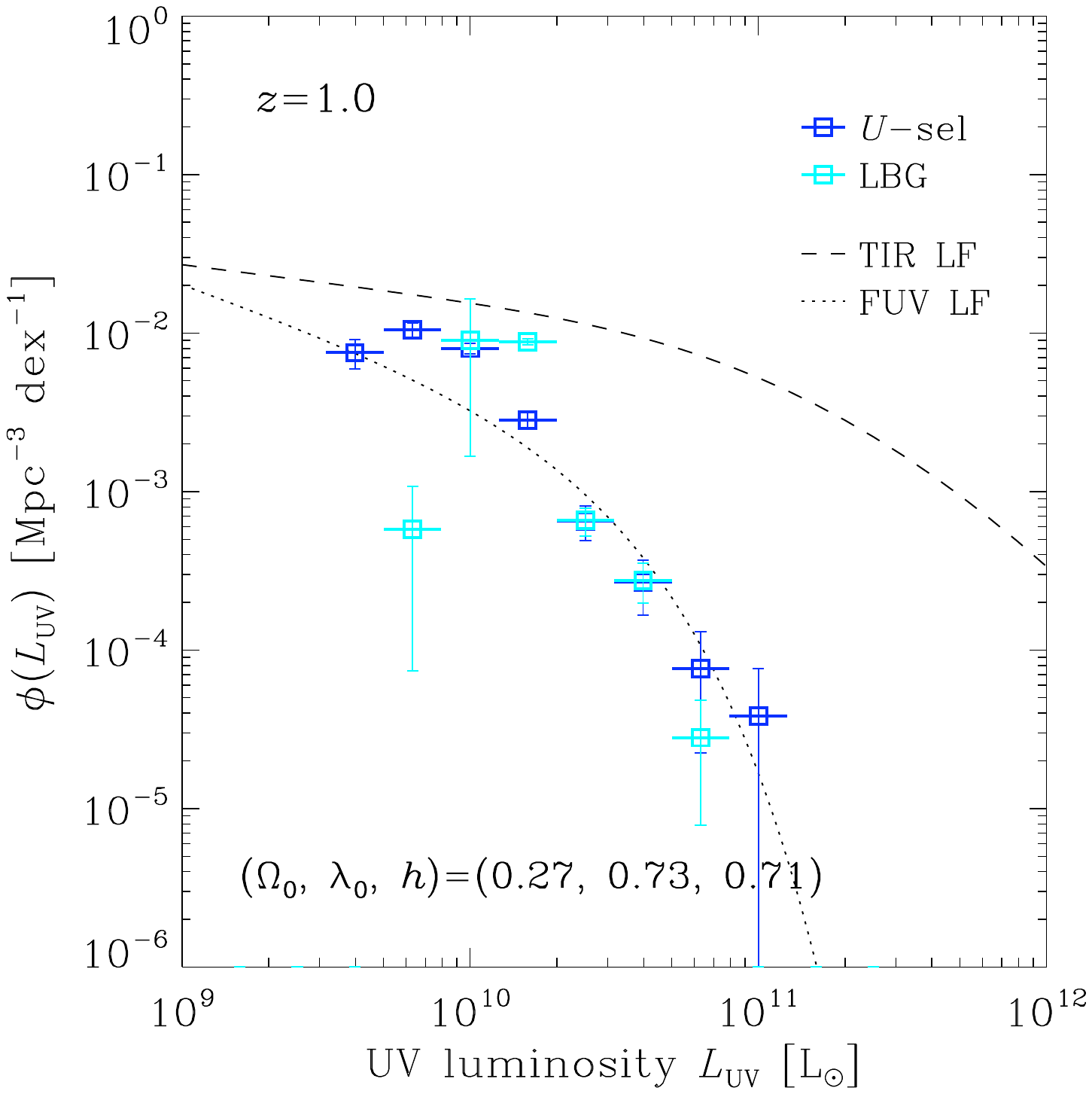}
  \caption{The UV (1600~\AA\ LFs of the sample at $z = 0.7$ and $z \simeq 1.0$
 ($U$-selected sample and LBGs)}
   \label{fig:uvlf}
\end{figure}

The most important but difficult point of this analysis is how to handle 
the two variables $L_{\rm UV}$ and $L_{\rm IR}$ at the same time.
We proceed a statistical analysis as follows:
\begin{description}
\item[Step 1] Since our sample is primarily selected at UV (GALEX FUV at  $z=0$  taken as reference ($\lambda = 
1530\;$\AA), GALEX NUV at $z = 0.7$ and EIS $U$-band at $z = 1$), 
we construct univariate UV LFs.
\item[Step 2] We bin the UV LFs, and estimate the distribution of the total 
IR luminosity estimated by equations~(\ref{eq:mir_tir12}) and 
(\ref{eq:mir_tir15}) at each bin.
\item[Step 3] We sum up the distribution functions of the total IR luminosity
along all the UV luminosity bins.
\end{description}

For Step~1, we used improved versions of two representative LF estimators
: $1/V_{\rm max}$-estimator \citep{schmidt1968} and $C^-$-estimator 
\citep{lynden_bell1971} in an optimal manner, explained and examined 
extensively by \citet{takeuchi2000a} and \citet{takeuchi2000b}.
We estimated the LFs of NUV-selected sample at $0.6 <  $z$.< 0.8$, $U$-band
selected sample at $0.8< z <1.2 $, and LBG samples $0.9 < z < 1.3$.
Since we are interested in the star-forming galaxies, we have omitted
known quasars/AGNs from our sample, as already explained in section 2.1.
These univariate UV LFs are estimated simply at their primary selection 
bands for  the NUV-selected $z = 0.7$ sample and $U$-band selected
$z = 1.0$ sample: we obtain LFs at $\sim 1400\;\mbox{\AA}$ and $\sim 1800\; \mbox{\AA}$ for 
$z=0.7$ and 1 respectively. 
As discussed in section 2.1, the NUV and  $U$ selected samples correspond to rest-frame wavelengths 
close enough to avoid K-corrections. It is not the case for the LBG sample primarily selected 
in NUV which corresponds to  $\sim 1100$~\AA\ rest-frame. We will go back to this issue 
in Section~\ref{subsec:uvlf}.

In Step~2, we should be careful for a significant number of upper
limits at MIPS $24\;\mu$m.
For this, we made use of the Kaplan-Meier estimator which enables us to
make use of the information content carried by the upper limits, originally
developed in the field of lifetime data analysis \citep{kaplan1958}.
Another desirable property of the Kaplan-Meier estimator is that we can
obtain its variance in an analytic form.
Formulations, derivations and some important properties will be discussed 
and explained elsewhere (Takeuchi et al. 2009, in preparation).

In Step~3, statistical errors are summed up in quadrature, i.e., the
variance from the primary univariate UV LF and that from the IR luminosity 
distributions. 
However, we did not include the variance caused by the density 
inhomogeneity of galaxies (often referred to as the cosmic variance).
We will discuss this issue in Section~\ref{subsec:uvlf}.

\subsection{UV luminosity functions}\label{subsec:uvlf}

We show the UV LFs for the samples at $z = 0.7$ and 1 in 
Figure~\ref{fig:uvlf}.
Since both $1/V_{\rm max}$ and $C^-$-estimates agree very well
with each other, we only show the latter in this paper.
We also estimated the UV LF of LBGs, but it is not obtained by
the same method as the other two, since the primary selection is done 
at NUV (for the redshift range of $0.9 < z < 1.3$, this corresponds to 
1100~\AA).
Then, we must use a bivariate method exactly the same as described 
in above Step~2: first we construct a univariate UV LF at 1100~\AA,
and construct distributions of UV luminosities at each bin from the $U$-band data 
($\sim 1700$~\AA\ at the rest-frame), and sum them
up along the UV luminosity bin.
The obtained UV LF of LBGs is presented in cyan symbols
in Figure~\ref{fig:uvlf}.

Roughly speaking, we observe that the UV LFs at redshift ranges of $z = 0.7$ and 
$z = 1.0$ agree with those of \citet{arnouts05}. 
This shows that our selection is appropriate for the purpose of this
study.
We may see that there are some discrepancies between the shapes of our UV LFs and those of \citet{arnouts05}.
The difference may be attributed to the difference in selection of galaxies, e.g., \citet{arnouts05} 
performed a NUV selection and $K$-corrected the flux, unlike our selection at $U$-band at $z=1$
without $K$-correction.
The different way photometry has been performed can also be at  the origin of a subtle difference: the GALEX deep 
fields are known to be crowded and down to $\mbox{NUV}=23$ mag, Sextractor does not separate 
accurately the sources leading to an under-density and a brightening of the sources. 
In the $U$-band the PSF is smaller, so less or not affected by confusion.

We then focus on the difference in the UV LF from $U$-band and the LBG LF.
As mentioned above, the former was constructed with the $U$-band selection 
with the univariate method, while the latter was constructed by the bivariate method: 
 we first select galaxies at NUV and with a  FUV-NUV colour criterion, then we  estimate 
the $U$-band luminosity distribution.
Therefore  LBG selection criteria at $z = 1.1$ are not very different from the $U$-selection at 
similar redshift ($z=1$) since both are based on a UV rest-frame selection. Nevertheless, in the LBG selection we miss 
UV-faint galaxies at $L_{\rm UV} < 10^{10} \;L_\odot$ when compared to the $U$-selection probably because of the combined effects of a selection at a shorter rest-frame wavelength for the LBGs and of a FUV-NUV criterion which selects only blue objects, as explained in section 2.2.
This suggests that the LBG selection criterion is more restricted to pick up UV-luminous galaxies.
We will come back to this point when we discuss the difference of the ${\rm UV} + {\rm IR}$ LF.

\subsection{The total UV$+$IR luminosity functions}

\begin{figure}[h]
\centering\includegraphics[width=9cm]{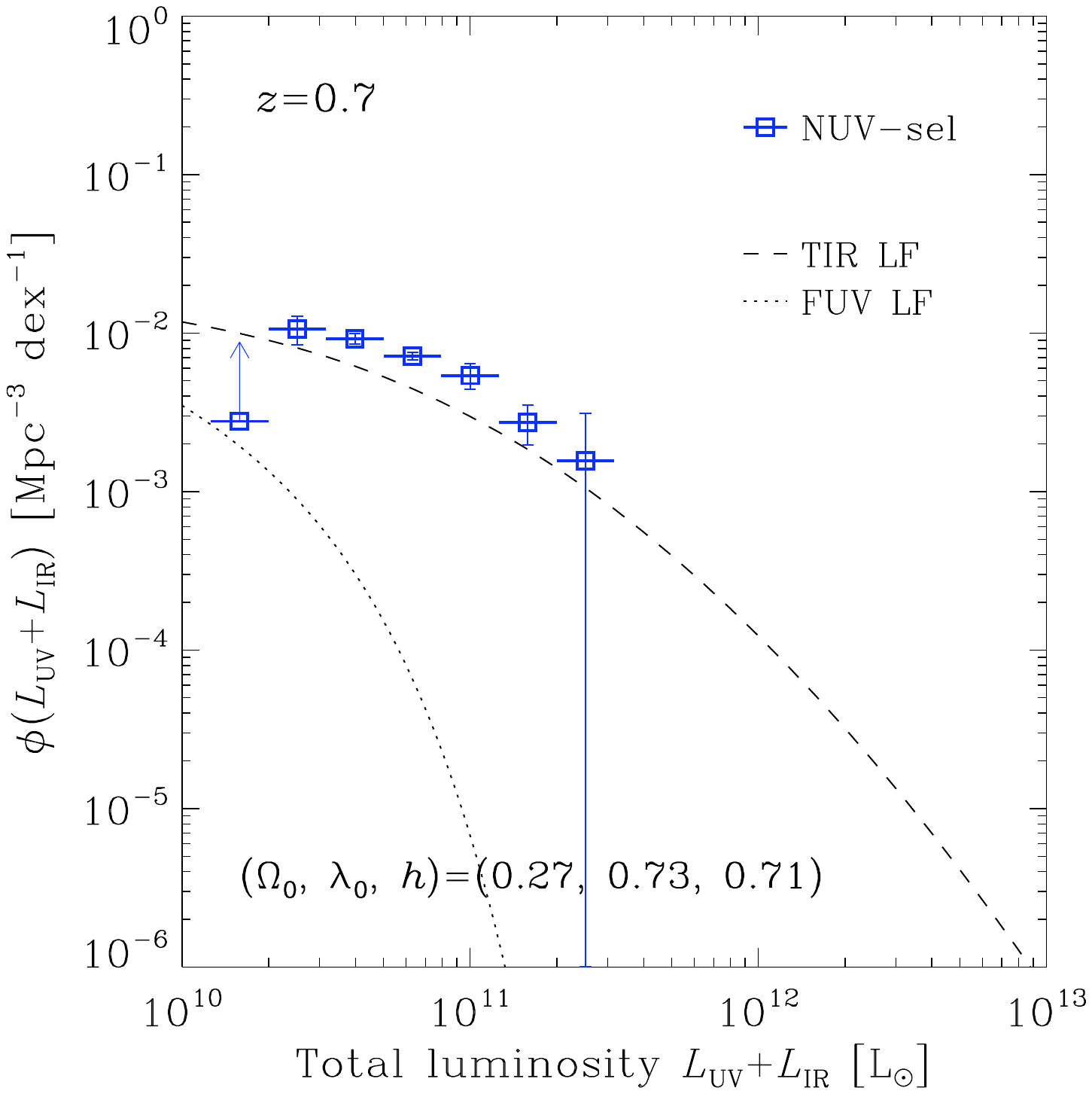}
\centering\includegraphics[width=9cm]{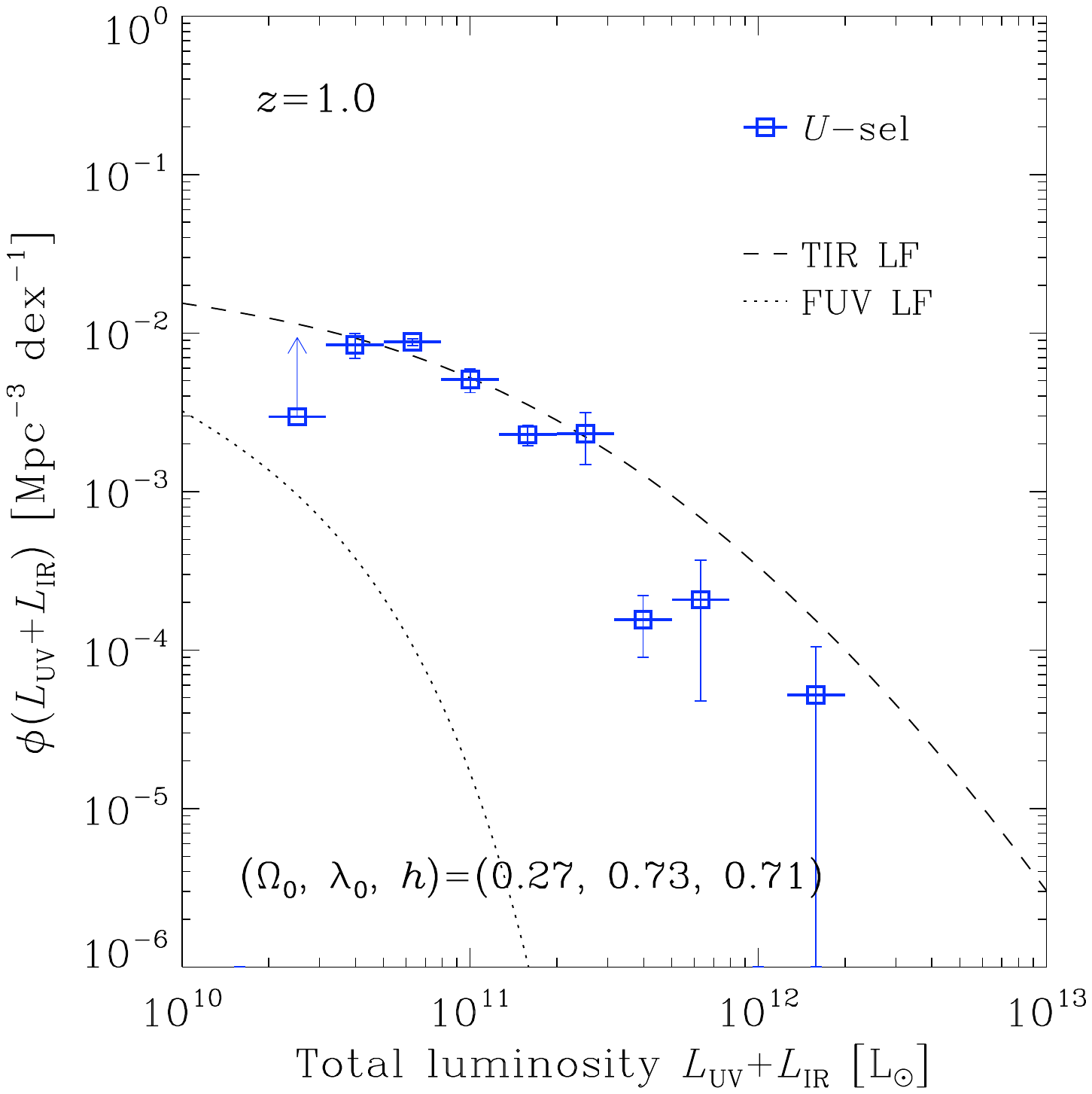}
\caption{The total $L_{\rm UV} + L_{\rm IR}$ luminosity functions at
$z = 0.7$ and 1.
We also show univariate UV LFs constructed from 
purely UV-selected samples  by \citet{arnouts05} (dotted lines), 
as well as univariate IR LFs made from purely IR-selected samples at
$24\;\mu$m by \citet{lefloch} (dashed lines).
The symbols are the LFs derived from our sample. 
The indicated errors are $1\sigma$ (68\% CL).
}\label{fig:blf}
\end{figure}

Here, we show the total UV+IR  LFs from our UV-selected
samples.
Again we stress that we addressed the upper limits of the sample at
MIPS $24\;\mu$m by Kaplan-Meier method, i.e., we have made a maximal
use of the observed information from IR.
We show the total LFs in Figure~\ref{fig:blf}.
Top panel shows the $L_{\rm UV} + L_{\rm IR}$ LF at $z = 0.7$, while
bottom panel is the one at $z = 1.0$.
In Figure~\ref{fig:blf}, we also show univariate UV LFs constructed from 
purely UV-selected samples by \citet{arnouts05} (dotted lines), 
as well as univariate IR LFs made from purely IR-selected samples at
$24\;\mu$m by \citet{lefloch} (dashed lines).
The symbols are the LFs derived from our sample. 
Errors are calculated analytically by the asymptotic variance formula
of the Kaplan-Meier estimator, convolved with the statistical error
of the univariate LFs at UV.
The indicated errors are $1\sigma$ (68~\% CL).
Because of the known limitation of the Kaplan-Meier estimator, the lowest
luminosity bins are underestimated (indicated by arrows on the symbols).

Clearly, the total LFs are much higher than the univariate UV LFs.
This means that most of the luminosity of a galaxy at these redshifts
is emitted in the IR.
Since the luminosity related to their star formation activity tends to
be emitted in the IR wavelengths \citep[e.g.,][]{ttt06}, the resulting
total LFs is consistent with this known fact.

 At $z = 0.7$, the total LF is even higher than the IR LF of \citet{lefloch},
but within a range of the cosmic variance \citep[$\sim 60$~\% for GOODS: ][]{somerville2004}.
Apart from this, it is rather consistent with the IR LF.
This is an expected higher-$z$ counterpart of the result discussed by
\citet{buat07-1} at $z = 0.0$.

 In contrast, the total LF is significantly lower than the IR LF at 
$z = 1.0$  for galaxies more luminous  than $\simeq 2 \times 10^{11} L_\odot$.
It is worth mentioning that the primary UV LF has an excess in
normalisation compared with the global univariate UV LF at the
same redshift.
Then, this deficiency of galaxies turns out to be quite significant.
Though the most luminous bin is disturbed by the symbol with a very
large error, we see a trend that the more luminous galaxies are, the
larger the discrepancy becomes.
This is a clear piece of evidence that our UV-selection misses intrinsically
luminous galaxies which are active in star formation. Such galaxies must be studied from an IR selection.
\citet{buat07-1} studied both IR and UV selections in the nearby universe and found that 
the more luminous galaxies (in terms of total $L_{\rm UV} + L_{\rm IR}$ luminosity) are present 
in the IR selection and suffer a very strong  dust attenuation. 
The galaxies more luminous than $\simeq 2 \times 10^{11} L_\odot$ exhibit a mean 
$\log(L_{\rm IR}/L_{\rm UV}) \ga 2$.  
The relation found between $L_{\rm UV} + L_{\rm IR}$ and $L_{\rm IR} / L_{\rm UV}$  at $z=0$ 
has been found to be globally  valid at higher $z$ for IR selections \citep{choi06,xu06,zheng07}.  
An $L_{\rm IR}/L_{\rm UV}$ ratio larger than 100 for galaxies more luminous  than 
$\simeq 2 \times 10^{11} L_\odot$ implies  that these galaxies are not detected at $z=1$ in our 
$U$ selection (limited to $\log(L_{\rm UV} [L_\odot]) >9.9 $)

At $z=0.7$ our UV selection goes deeper ( $\log(L_{\rm UV} [L_\odot]) > 9.3$), therefore galaxies with 
a larger dust attenuation can be detected in UV.  
It is also interesting to note that the galaxies with the highest $L_{\rm IR} / L_{\rm UV}$ are found at 
$z = 0.7$ (Fig.~\ref{plotluv}) for the UV faintest galaxies and that the fraction of galaxies with 
$\log(L_{\rm IR} / L_{\rm UV})>1$ is  globally higher at $z=0.7$ than at any other redshift 
(Fig.~\ref{pourcents}).
However, it is puzzling to invoke a drastic evolution from $z = 0.7$ to $z = 1.0$. 
Since the cosmic time differs by less than a few Gyr, such an evolution should be very fast.

As a conclusion, up to $z=1$ UV rest-frame observations must be much deeper (by more than 5 mag) 
than the expected limit in bolometric luminosities if one must be able to detect most of the 
star forming galaxies.

\begin{figure}[h]
\centering\includegraphics[width=10cm]{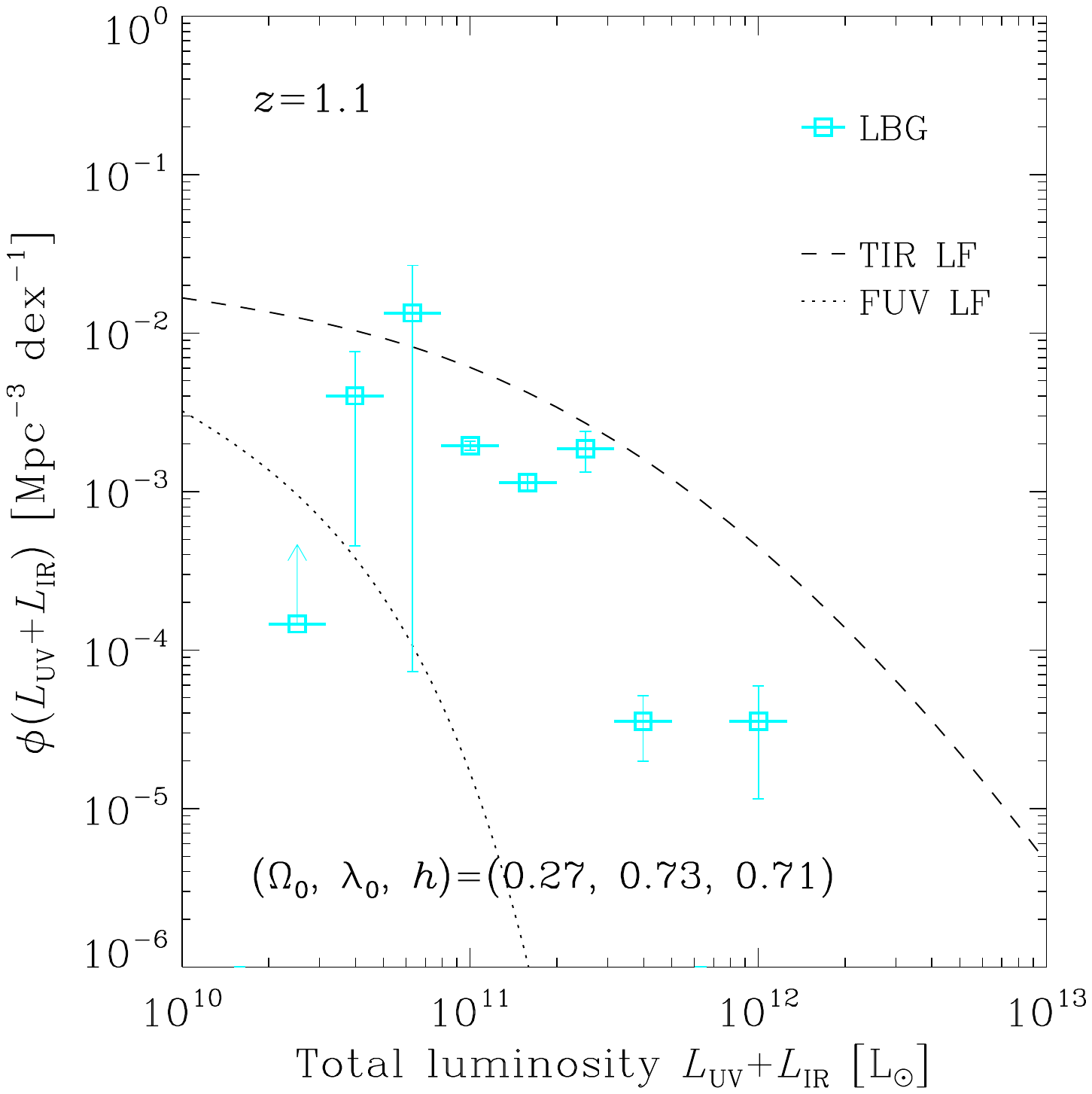}
\caption{
The total $L_{\rm UV} + L_{\rm IR}$ LF of the LBG sample.
The open squares represent the total LF of LBGs.
Other symbols are the same as in Figure~\ref{fig:blf}.
}\label{fig:blf_lbg}
\end{figure}

Figure~\ref{fig:blf_lbg} shows the total UV+IR LF of the LBG sample.
The deficiency of total LF is more prominently seen in the LBG LF.
In this case, the deficiency ranges toward lower luminosities
$\simeq 4 \times 10^{10} L_\odot$.
Considering the LBG sample selection which makes use of NUV and FUV fluxes observed at $2310$ and $1530$~\AA~
by GALEX, this trend may be understood consistently:
the LBG sample consists of galaxies with less 
extinction on average, leading to less IR luminosities with 
respect to the same $L_{\rm UV}$ as shown in section 3. 
The deficiency with respect to the $U$-selection affects  the faintest bins of the LBG LF  (Fig.~\ref{fig:uvlf} and discussion in section 5.2).
Since the dispersion of the $L_{\rm IR}/L_{\rm UV}$  distribution is very large as we have seen before, the contribution of these bins to 
the number density of galaxies is significant and  the deficiency of 
galaxies affects all the range of the total ${\rm UV} + {\rm IR}$ LF.

\subsection{Discussion}

As discussed by many authors \citep[e.g.,][]{chary01,ttt,caputi07,rieke08},
the monochromatic MIR luminosity-to-total IR luminosity conversion 
plays an important role, especially because of the currently limited
number of deep multi-band observations at FIR.
Since the intrinsic scatter in the linear regression is not very small,
these ``calibration formulae'' inevitably have significant uncertainty.
Then, it makes sense to examine how different formulae change our results,
especially the deficiency of intrinsically luminous galaxies in the 
UV-selection.

\begin{figure}[h]
 \centering\includegraphics[width=10cm]{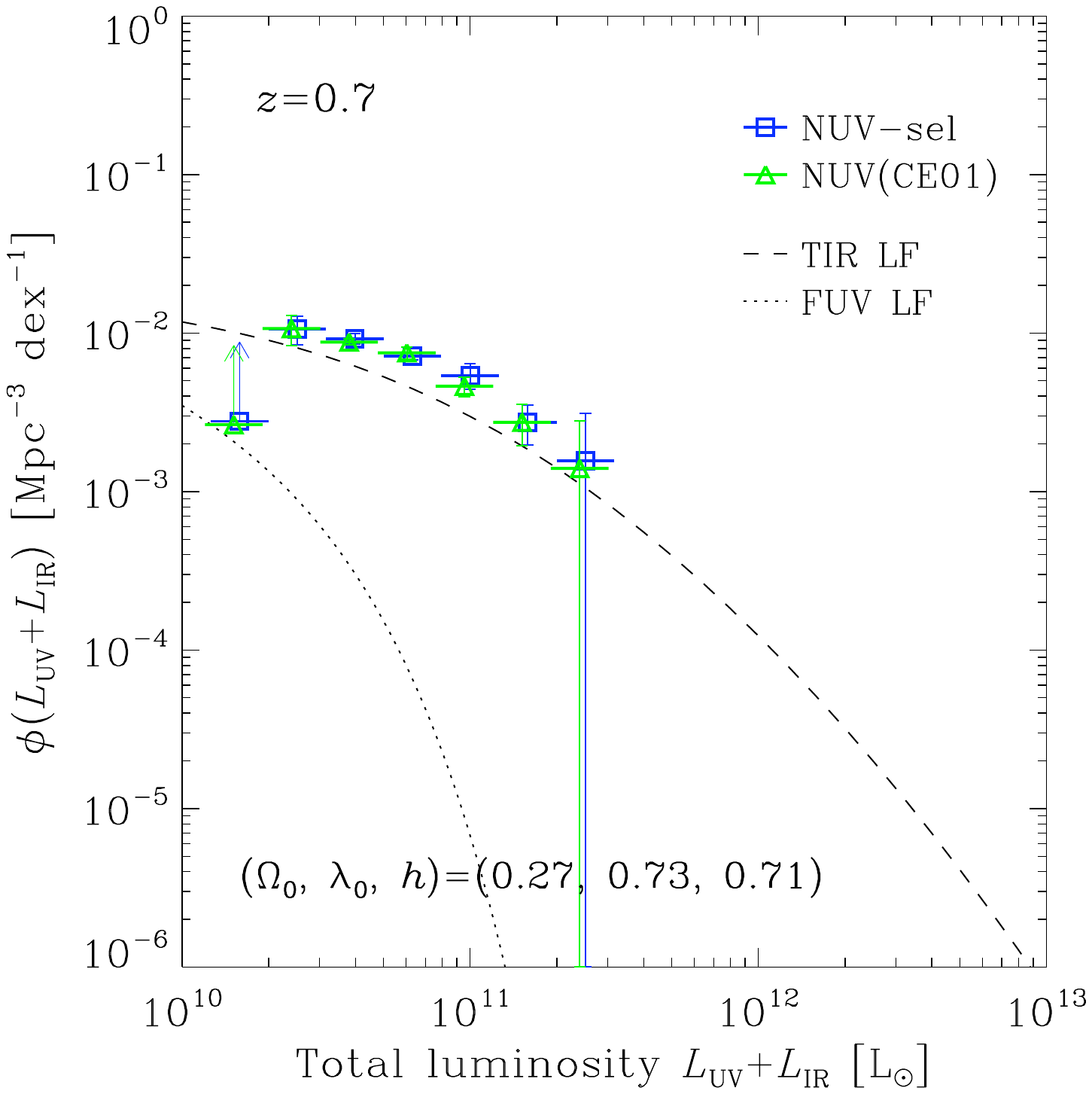}
 \centering\includegraphics[width=10cm]{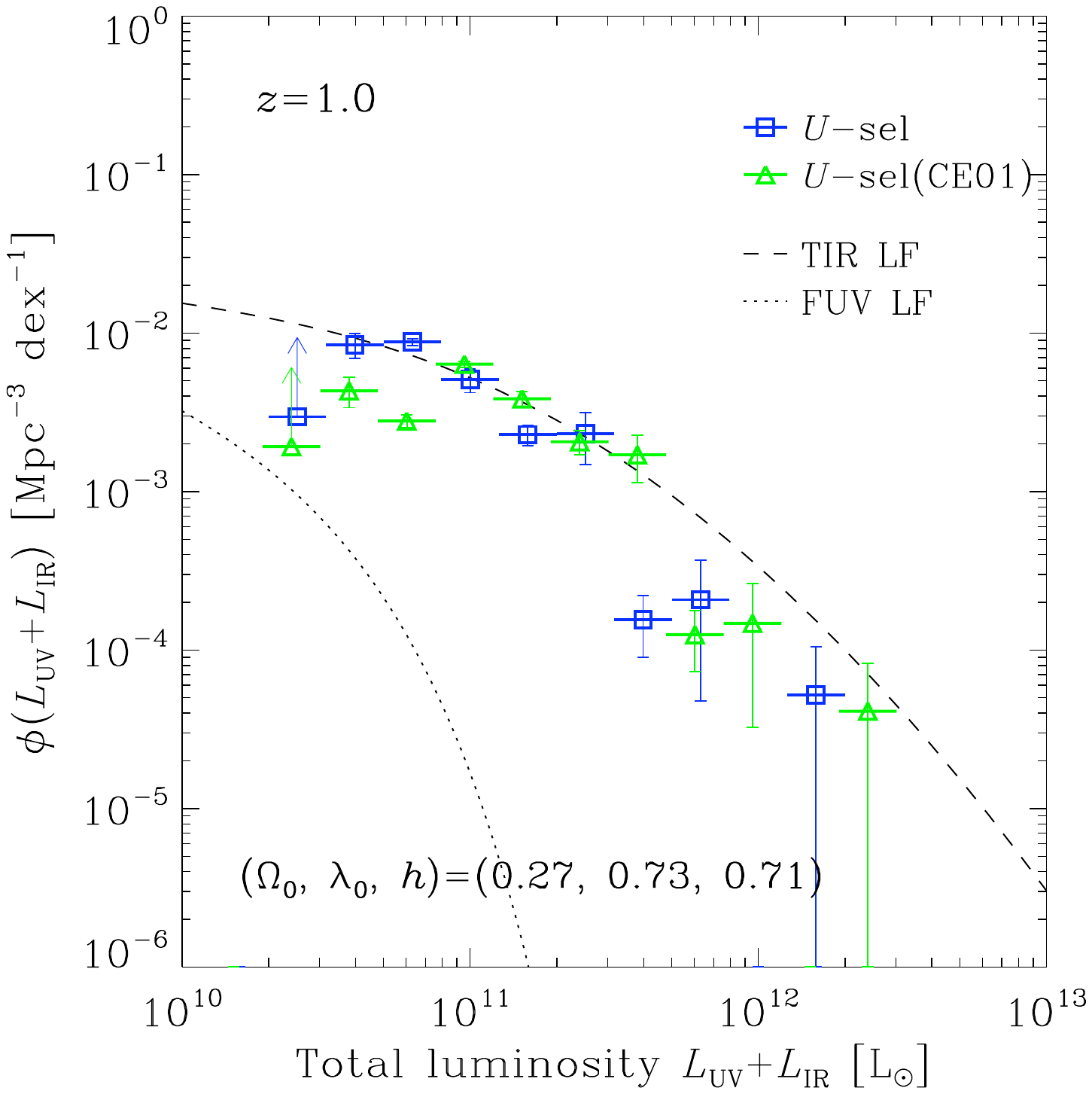}
\caption{Comparison between the total LFs with the $L_{\rm IR}$ estimator
of \citet{ttt} and those with \citet{chary01}.
}\label{fig:blf_comp}
\end{figure}

{}To test this, we have estimated the total ${\rm UV} + {\rm IR}$
LFs exactly in the same manner but with CE01 conversion.
The resulting LFs are shown in Figure~\ref{fig:blf_comp}.
In Figure~\ref{fig:blf_comp}, open squares represent the LFs with
the formula of \citet{ttt}, while open triangles are the ones with
CE01 conversion.
All the other symbols are the same as in Figure~\ref{fig:blf}.

At $z = 0.7$, since the difference of these formulae is quite small
(cf.\ upper panel in Fig.~\ref{calib-lir}), the results are almost the same.
At $z = 1.0$, the difference is visible between the two estimates.
As we have seen in Figure~\ref{calib-lir}, CE01 formula gives larger
IR luminosity.
Hence, it produces larger total luminosity in $L_{\rm UV}+L_{\rm IR}$ 
with respect to the same $L_{\rm UV}$.
As a result, the discrepancy between the total LF and the IR LF becomes
smaller, but still statistically significant.

At $z\simeq 2$ and $\simeq 3$, \citet{reddy08} were  able to reproduce all the 
IR LF up to $L_{\rm IR} = 10^{12}~L_\odot$ from only UV-optical data with even an excess of faint sources as compared to the results from IR surveys alone. They built the UV rest-frame LF with Monte Carlo simulations to recover all the star forming galaxies; then, to recover the IR LF they assumed either a constant dust attenuation distribution irrespective of UV luminosity or a decrease of the average dust attenuation for UV faint galaxies.
 Conversely, at $z \simeq 1$ dust attenuation is found not very dependent on  UV luminosity with only a slight decrease for UV luminous galaxies.  
We cannot reconstruct the bright end of the IR LF from a UV-selected sample. The IR luminous galaxies, observed in IR surveys, exhibit a very large dust attenuation which makes them undetected in UV (rest-frame). 
Since we use actually observed IR and UV flux densities (including upper limits) our method can be 
considered as being secure, although it is dependent on the validity of the MIR to total IR luminosity 
conversion. 
The method of Reddy et al.\ also suffers from the uncertainty in the MIR to total IR luminosity 
conversion which is particularly large at $8\;\mu$m rest-frame and on the accuracy of 
dust attenuation factors estimated from the UV-optical alone (see discussion in section 3.2).
Nevertheless if we trust both results it implies a lower fraction of galaxies intrinsically UV$+$IR luminous and with a large dust attenuation at  $z\simeq 2-3$ than at $z\simeq 1$.
We will re-investigate this issue by  using IR-selected
samples up to $z=1$ in a fully bivariate manner 
(Takeuchi et al. 2009, in preparation). 
The future observations of HERSCHEL should give us the high redshift IR selected samples 
necessary to solve this question.

\section{Conclusions}
We have analysed the IR emission of galaxies selected in UV rest-frame from $z = 0$ to $z=1$. 
The samples were built to be very homogeneously selected in terms of wavelength and luminosities. 
We also  considered a sample of Lyman Break Galaxies at $z\simeq 1$.
\begin{enumerate}
\item The $L_{\rm IR}/L_{\rm UV}$ ratio was used as a proxy for dust attenuation. 
For the bulk of our galaxy samples, this dust attenuation  is found  not to vary significantly with  $z$ and 
less than 20~\% of the sample galaxies have a $L_{\rm IR}/L_{\rm UV}>10$. 
Anyway, some evolution is seen in the extreme regimes of high and low $L_{\rm IR}/L_{\rm UV}$ ratio. 
The most luminous UV  objects  ($L_{\rm UV} \simeq 2 \times 10^{10} L_\odot$) are only present at $z=1$ and 
exhibit a very low dust attenuation. When $L_{\rm UV} \la 2 \times 10^{10} L_\odot$, 
the fraction of galaxies  with a high $L_{\rm IR}/L_{\rm UV}$ is larger at $z>0$  than in the nearby 
universe and the galaxies with the largest dust attenuation are the faintest ones in our samples  
($L_{\rm UV} \simeq 3 \times 10^{9} L_\odot$) . 
However these results all depend on the MIR-total IR calibration which is uncertain. 
Dust attenuation increases with the $K$ luminosity in a similar way at all redshifts.  
A residual trend is found with the UV luminosity:  when $L_{\rm UV}$ increases, galaxies of 
a given $L_K$ have a lower $L_{\rm IR}/L_{\rm UV}$. A relation between  $L_{\rm IR}/L_{\rm UV}$,  
$L_{\rm UV}$ and $L_K$  is given. 
LBGs at $z = 1$ seem to be less extinguished than UV selected galaxies of similar UV luminosity 
and at same $z$. 
Since the UV luminosity of galaxies globally increases with $z$,  these trends found with the UV 
luminosity must be accounted for to interpret the evolution with redshift  of 
$L_{\rm IR}/L_{\rm UV}$ reported in previous studies.

\item Massive and UV luminous galaxies ($ \log( M_{\rm star})> 10.8 (M_{\odot})$  and $\log L_{\rm UV} >10.4 (L_\odot)$) are  found very active in star formation (large  SSFR) whereas  fainter galaxies of similar mass 
exhibit a larger range of SSFR.  LBGs and UV selected galaxies have similar SSFR.
  
\item We constructed LFs with total luminosity related to star 
formation activity, $L_{\rm UV} + L_{\rm IR}$ from our UV-selected galaxy samples.
We have used the Kaplan-Meier estimator to make use of information carried
by IR detections and upper limits in a coherent manner.
The resulting total ${\rm UV} + {\rm IR}$ LFs are much higher than
the univariate UV LFs from purely UV-selected samples.
This means that most of the luminosity produced by star formation 
activity is emitted in the IR wavelength range.
Though at $z = 0.7$, the total LF we obtain is consistent (even higher because of
a density excess) with the univariate IR LF, we find a clear deficiency 
of galaxies in the total LF at $z = 1.0$ for galaxies more luminous than $\simeq 2 \times 10^{11} L_\odot$
This result is not significantly affected by different total IR luminosity calibration formula.
Thus, we conclude that the IR LF cannot be reconstructed solely from our 
UV-selected galaxies at $z=1$ and that deeper data are needed in order to detect galaxies 
with a large  $L_{\rm IR}/L_{\rm UV} $. Practically,  to detect most of the star forming galaxies down to a given bolometric magnitude, UV rest-frame observations must be deeper  than this bolometric limit by at least 5 mag (corresponding to  ($L_{\rm IR}/L_{\rm UV} \simeq 100$).
The deficiency in the total LF is found much higher for the LBG selection affecting the whole range of 
luminosity explored in this work for these objects (i.e. $\ga 4 \times 10^{10} L_\odot$).

\end{enumerate}

\begin{acknowledgements}
TTT has been supported by Program for Improvement of Research 
Environment for Young Researchers from Special Coordination Funds for 
Promoting Science and Technology, and the Grant-in-Aid for the Scientific 
Research Fund (20740105) commissioned by the Ministry of Education, Culture, 
Sports, Science and Technology (MEXT) of Japan.
We thank Akio K.\ Inoue and Hiroyuki Hirashita for fruitful discussions.
TTT and KLM are partially supported from the Grand-in-Aid for the Global 
COE Program ``Quest for Fundamental Principles in the Universe: from 
Particles to the Solar System and the Cosmos'' from the MEXT.
\end{acknowledgements}

\end{document}